\documentclass[10pt,tightenlines,onecolumn,amssymb, nobibnotes, aps, prd]{revtex4-2}
\usepackage[paperwidth=210mm,paperheight=297mm,centering,hmargin=2cm,vmargin=2.5cm]{geometry}
\setcitestyle{authoryear,round}
\usepackage[english]{babel}
\usepackage{graphicx}
\usepackage{xcolor}
\graphicspath{{./Figures/}}

\usepackage{amsmath}
\usepackage{xfrac}
\definecolor{grALGO}{RGB}{230,230,230}
\usepackage{bm}
\usepackage{hyperref}
\hypersetup{
    bookmarks=true,         
    unicode=false,          
    pdftoolbar=true,        
    pdfmenubar=true,        
    pdffitwindow=false,     
    pdfstartview={FitH},    
    pdftitle={Multiscale modeling of inelastic materials with Thermodynamics-based Artificial Neural Networks (TANN)},    
    pdfauthor={F. Masi, I. Stefanou},     
    pdfnewwindow=true,      
    colorlinks=true,       
    linkcolor=blue,          
    citecolor=blue,        
    filecolor=blue,         
    urlcolor=blue        
}
\usepackage[font=small,labelfont=bf]{caption}
\usepackage{subcaption}
\usepackage[algoruled]{algorithm2e}

\begin{document}
\title{\Large \textbf{Multiscale modeling of inelastic materials with Thermodynamics-based Artificial Neural Networks (TANN)}}

\author{Filippo Masi\footnote{Corresponding author}}
 \email{filippo.masi@ec-nantes.fr}
\author{Ioannis Stefanou}
 \email{ioannis.stefanou@ec-nantes.fr}
\affiliation{Nantes Université, Ecole Centrale Nantes, CNRS, GeM, UMR 6183, F-44000 Nantes, France.
}%

\date{\today}

\begin{abstract}
\noindent The mechanical behavior of inelastic materials with microstructure is very complex and hard to grasp with heuristic, empirical constitutive models. For this purpose, multiscale, homogenization approaches are often used for performing reliable, accurate predictions of the macroscopic mechanical behavior of solids and structures. Nevertheless, the calculation cost of such approaches is extremely high and prohibitive for real-scale applications involving inelastic materials. Recently, data-driven approaches based on machine learning and, in particular, deep learning have risen as a promising alternative to replace ad-hoc constitutive laws and speed-up multiscale numerical methods. However, such approaches require huge amounts of high quality data, fail to give reliable predictions outside the training range and lack a rigorous frame based on the laws of physics and thermodynamics. As a result, their application to model materials with complex microstructure in inelasticity is not yet established.

\noindent Here, we propose the so-called Thermodynamics-based Artificial Neural Networks (TANN) for the constitutive modeling of materials with inelastic and complex microstructure. Our approach integrates thermodynamics-aware dimensionality reduction techniques and thermodynamics-based deep neural networks to identify, in an autonomous way, the constitutive laws and discover the internal state variables of complex inelastic materials. The ability of TANN in delivering high-fidelity, physically consistent predictions is demonstrated through several examples both at the microscopic and macroscopic scale. \\
The efficiency and accuracy of TANN in predicting the average and local stress-strain response, the free-energy and the dissipation rate is demonstrated for both regular and perturbed two- and three-dimensional lattice microstructures in inelasticity. TANN manage to identify the internal state variables that characterize the inelastic deformation of the complex microstructural fields. These internal state variables are then used to reconstruct the microdeformation fields of the microstructure at a given state. Finally, a double-scale homogenization scheme (FEM$\times$TANN) is used to solve a large scale boundary value problem. The high performance of the homogenized model using TANN is illustrated through detailed comparisons with microstructural calculations at large scale. An excellent agreement is shown for a variety of monotonous and cyclic stress-strain paths.

\end{abstract}

\maketitle

\section{Introduction}
\noindent Many problems in science and engineering involve complex materials, that are heterogeneous and multiscale in nature (e.g. geo-materials, bio-materials, meta-materials). The mechanical response of such systems is complex and hard to grasp with heuristic, empirical constitutive models.

\indent For this purpose, multiscale numerical modeling is frequently used. Multiscale  approaches allow to capture in detail the effects of the microstructure and its influence at the macroscopic scale. This is accomplished by iteratively solving a boundary value problem (BVP) of a representative cell of the microstructure under specific boundary conditions (auxiliary problem). Then, based on homogenization, the intrinsic constitutive behavior of the microstructure is upscaled to the macroscopic level  \cite[see][]{Bakhvalov1989,geers2010multi}. These techniques lead to mixed numerical schemes such as the FE$^2$ method \citep{Feyel2003,lloberas2019reduced, Eijnden2016} and FEM$\times$DEM \citep{Nitka2011, Nguyen2014}, depending on the nature of the microstructure. However, the computational cost of these methods is extremely high for real-scale applications, if not prohibitive. The reason is that the numerical solution of the auxiliary problem that needs to be calculated millions of times is computationally expensive.

In the last decades, machine learning has been widely employed in order to speed-up the aforementioned numerical methods. Artificial Neural Networks (ANN) are one of the most promising models. Recent examples have shown that ANN can successfully encapsulate several aspects of the constitutive behavior of the underlying microstructure and provide the necessary information to the macroscopic scale with reduced calculation cost \citep[see][for a review]{ghaboussi1991,lefik2003artificial,boso2012multiscale,mozaffar2019deep,liu2019exploring,huang2020learning,vlassis2021sobolev,liu2021review,mianroodi2021teaching,zhou2021learning}. Yet, deep learning-based constitutive modeling has some major weaknesses. Among the most significant, we note (a) the limited extrapolation capabilities for predicted values far beyond the training range; (b) the need of vast amounts of high quality data (i.e. with reduced noise); and, more important (c) the lack of a framework based on physics as trained ANN do not necessarily respect basic principles of physics. 

One way to tackle the above weaknesses is to design ANN architectures that respect, by construction, the laws of physics \citep{raissi2019physics} and thermodynamics \citep{masi2021thermodynamics,hernandez2022thermodynamics}. This was recently achieved for homogeneous inelastic materials \citep[Thermodynamics-based Artificial Neural Networks -- TANN --][]{masi2021thermodynamics, masiTANNspigl}. In that work, the constraints of thermodynamics were imposed through the computation of the numerical derivatives of the network with respect to its inputs. In particular, the derivatives of the free-energy and their relation with the stress, the dissipation rate, and the internal state variables were hardwired in the architecture of the neural network. This lead to accurate and robust predictions even for unseen data with noise. However, this approach does not cover the upscaling of heterogeneous, complex materials as it needs the a priori identification of the internal state variables.

\noindent Learning the constitutive response of inelastic materials with microstructure is challenging. The bulk of the existing methods in the literature makes use of the deformation history of the microstructure. Among the different existing approaches, we refer to the most recent ones: \citet{Wang2018,ghavamian2019accelerating,zhang2020machine,sun2021data,PLED2021113540,gartner2021,li2021physics,sorini2021convolutional,SAHA2021113452}.
Nevertheless, physics constraints are not considered in these methods \citep[see][for a comprehensive review]{peng2021multiscale}.\\
There exist only few works where physics-aware neural networks are used to speed-up multiscale simulations \citep[see ][among others]{PhysRevE.102.053304,HALL2021110192,hernandez2021deep,LI2021100429,yin2022interfacing}. \citet{PhysRevE.102.053304} developed a data-driven approach based on processing histories to satisfy physically correct boundary conditions of multiscale systems. \citet{HALL2021110192} used graph-informed neural networks to recover reduced-order models for physics-based representations of multiscale systems. With the same target of deriving surrogate models for multiscale systems, \citet{hernandez2021deep} developed structure preserving neural networks, which enforce thermodynamics restrictions through the GENERIC framework \citep{grmela1997dynamics}.

Here we propose an alternative method, which is based on the discovery of a reduced set of Internal State Variables (ISV) at the level of the microstructure. This allows us to separate the solution of the macroscale boundary value problem from the one of the microstructure.  For this purpose, we resort to the theory of internal state variables \citep{coleman1967thermodynamics,maugin1994thermodynamicsA} to efficiently and accurately account for dissipative processes, without the knowledge of the stress-strain history of the microstructure.\\
The internal state variables are identified using thermodynamics-based dimensionality reduction techniques. In the applications herein considered, we opt for autoencoders, which identify the necessary internal state variables from the \textit{internal coordinates} of the microstructure (e.g. displacement, velocity fields). Other model-order reduction techniques may be preferred \citep{brunton2022data}.\\
The autoencoder is combined with the architecture of TANN in order to yield a thermodynamically consistent material state space. Consequently, the high-dimensional space of the internal coordinates of the microstructure is compressed into the internal state variables. Reconstruction of micromechanical fields is possible at any given state, through the (trained) decoder network.

\noindent The proposed method is able, for the first time, to translate theoretical principles derived from thermodynamics (such as the identification of internal state variables) into artificial neural networks for the robust, accurate and thermodynamically consistent modeling of complex materials. Moreover, with this approach, data-sets do not need to be ordered in time. This allows data generation using random stress-strain paths, rather than specific stress-strain time histories. This is an advantage for performing large multiscale simulations and data generation.

\noindent Several benchmarks are presented to demonstrate the efficiency and accuracy of the proposed methodology. As an example we use two- and three-dimensional lattices with elastoplastic elements. Despite the apparent complexity of the microstructure, TANN manage to (a) identify a reduced space of internal state variables that characterize the microstructural fields and the average material behavior, (b) learn the free-energy and the dissipation rate of the microstructure, and (c) predict the average stress-strain response for a given strain increment. The trained network is then used in a Finite Element (FE) code (as a material) in order to perform large scale simulations. This multiscale approach is denoted here as FEM$\times$TANN to distinguish from FE\textsuperscript{2}, FEM$\times$DEM, and similar. The results are compared to detailed micromechanical computations at large scale. The comparison is made at two levels. At the macroscopic level, we compare the free-energy, dissipation and the stress-strain response. At the microscopic level, we compare the micromechanical fields of displacement and stress, which are reconstructed by decoding the internal state variables. In this double-scale computational approach, we use a double scale incremental asymptotic homogenization scheme of first order. The comparisons show an excellent agreement between the detailed micromechanical simulations and the homogenized TANN analyses for a variety of monotonous and cyclic stress-strain paths.\\

The paper is structured as follows. Section \ref{sec:theory} presents the theoretical background of the proposed methodology. In Section \ref{sec:tann} we briefly recall the previous framework of Thermodynamics-based Artificial Neural Networks (TANN), with a priori determination of ISV. The concept of internal coordinates is then discussed and a thermodynamics-aware dimensionality reduction technique is presented. These concepts are used to develop the new class of TANN without any a priori choice of the internal state variables for microstructured materials. Finally, Sections \ref{sec:app1} and \ref{sec:app2} present benchmarks of the proposed methodology for lattice materials, with inelastic microstructure. In particular, Section \ref{sec:app1} shows the successful implementation of TANN in deriving and modeling the material response of two-dimensional lattice cells of different geometry and complexity. Section \ref{sec:app2} focuses on FE multiscale simulations of large scale three-dimensional lattice structures, relying on TANN. Several loading scenarios, of increasing complexity, are investigated and the accuracy of the proposed methodology is illustrated. Finally, we discuss  the computational cost of our analyses, which is several orders of magnitude lower than the detailed three-dimensional micromechanical simulations.\\

\noindent The following notation is used throughout the manuscript: $a\cdot b = a_i b_i$, $a:b=a_{ij} b_{ij}$, $a\otimes b = a_i b_j$. Einstein's notation is implied for repeated indices.

\section{Thermodynamic considerations}
\label{sec:theory}

\subsection{General thermodynamic setting}
\noindent Consider a solid with reference configuration $\mathcal{B}$.
The energy balance in local form is written as follows
\begin{equation}
\dot{e} + \text{div}\,q= \wp^{\text{in}} +r,
\label{eq:thermo0}
\end{equation}
where $e$ and $\dot{e}$ are the volume density of the internal energy and its material time derivative, respectively; $\pi^{\text{in}}$ the power of internal forces; $q$ the heat flux; and $r$ external heat sources.\\
\indent An equivalent Cauchy continuum description is adopted here by assuming $\wp^{\text{in}} = p : \dot{f}$, with $p$ being the first Piola-Kirchhoff stress tensor and $\dot{f}$ the time derivative of the deformation gradient $f$.

\noindent The local form of the second law of thermodynamics reads
\begin{equation}
d^{*} = \dot{\eta} - \text{div}\,\left(\theta^{-1}q\right) -\theta^{-1}r\geq 0,
\label{eq:thermo1}
\end{equation}
where $d^{*}=d +d^{\text{th}}$ is the volume density of the total dissipation rate; $d$ and $d^{\text{th}}$ the mechanical and thermal dissipation rate, respectively; $\eta$ the volume density of entropy; and $\theta$ the absolute temperature.
Combining the first and second laws leads to the Clausius-Duhem inequality
\begin{equation}
d^{*} = p : \dot{f} - \dot{e} +\theta\dot{\eta} -\theta^{-1} q\cdot m\geq 0,
\label{eq:thermo2}
\end{equation}
where $m$ is the temperature gradient.
The last term in inequality (\ref{eq:thermo2}) coincides with the thermal dissipation rate, i.e., $d^{\text{th}}=-\theta^{-1} q\cdot m$. By assuming $d^{\text{th}}\geq 0$ and denoting the free-energy density with $\psi=e-\eta\theta$, the Clausius-Duhem inequality can be reformulated into
\begin{equation}
d =p:\dot{f} -\dot{\psi} -\dot{\theta}\eta \geq 0.
\label{eq:thermo3}
\end{equation}

\subsection{Volume averaging under periodic conditions}
\label{sec:volavg}
\noindent Let $\mathcal{V}\subset\mathcal{B}$ be an arbitrary volume and denote the average value of a quantity in $\mathcal{V}$ by 
\begin{equation}
\langle{\,\cdot\,}\rangle= \frac{1}{|\mathcal{V}|}\int_{\mathcal{V}} \cdot \,dx, \quad \text{ with } |\mathcal{V}|=\int_{\mathcal{V}} dx \neq 0.
\end{equation}
Assuming periodic displacement fields and anti-periodic traction vectors, we obtain a volume average expression of the first and second law of thermodynamics
\begin{equation}
D=S:\dot{F} - \dot{\Psi} - H^{\dagger}  \mathit{\dot{\Theta}},  \quad \text{ and } \quad D\geq 0,
\label{eq:thermo4}
\end{equation}
where we denoted $S\triangleq\langle\,p\rangle$; $F\triangleq\langle\,f\rangle$; $\Psi\triangleq\langle\psi\rangle$; $\Theta\triangleq\langle\theta\rangle$; $D\triangleq\langle\,d\rangle$; and $H^{\dagger}$ such that $\langle\eta \dot{\theta}\rangle=H^{\dagger} \dot{\Theta}$.\\
In the following we will work with the above volume averaged quantities, which, based on the Thermodynamic-based Artificial Neural Network proposed in Section \ref{sec:tann}, will allow us to perform double-scale asymptotic homogenization of heterogeneous, complex microstructures, such as the ones presented in Section \ref{sec:app2}.

\subsection{Constitutive description in terms of internal state variables}
\label{subsec:state}
\noindent Key point in defining thermodynamic restrictions for a material is to specify its state space, $\varsigma$, i.e., the set of variables its constitutive behavior depends on. In general, two possibilities exist.\\
\indent The first assumes that the present state of the material depends on the whole history \citep[rational thermodynamics, see][]{truesdellrational}. The state, at time $t$, is identified by $\varsigma(t)= \tilde{\varsigma}\big(\chi(t-t')\big)$, with $0\leq t'< t$, and $\chi$ are the so-called state variables, commonly employed in classical thermodynamics to study elastic materials \citep[e.g. temperature, strain measures, see][]{maugin1994thermodynamicsA}. It follows that the stress and all other dependent variables are functions of the history of the state variables.\\
\indent The second possibility consists in identifying the state space with internal state variables \citep{coleman1967thermodynamics}, defined at the current time $t$. In this case, the process history is accounted through the evolution in time of these internal state variables. The state of the material is univocally defined by (1) the state variables $\chi$, and (2) a certain number $N_{Z}$ of internal state variables (ISV) $Z=\{Z_k\}$ ($k=1,\dots,N_Z$), which are supposed to describe the internal material structure. It follows that the material state space is identified by $\varsigma(t) = \tilde{\varsigma}\big(\chi(t),Z(t)\big)$, for all $t$. 
The introduction of internal state variables renders possible the use of large state spaces, i.e., the constitutive relations can be described by maps, local in time, thus avoiding the use of their histories which appear in small state spaces \citep{maugin1994thermodynamicsA}. From the physics point of view, internal state variables are introduced to compensate for the lack of a precise description of microscopic phenomena (e.g. instabilities and rearrangements), which manifest themselves as irreversibilities at the macroscopic scale \citep{muschik}. Internal variables are macroscopic, measurable quantities, but not controllable, i.e., they cannot a priori be adjusted to a prescribed value through a direct action on the system. For this reason, they are also called in the literature `hidden' variables \citep{mcdowell2005internal}.

\noindent Here we follow this second approach and we write
\begin{equation*}
\begin{split}
\Psi &= \hat{\Psi}\left(\Theta,F,Z\right),
\end{split}
\end{equation*}
By noticing that the time differentiation and the volume average operators commute (Lagrangian representation), the time derivative of the Helmholtz free-energy $\Psi$ can be expressed as
\begin{equation}
\dot{\Psi} = \dfrac{\partial\Psi}{\partial \Theta} \,\dot{\Theta} + \dfrac{\partial \Psi}{\partial F} : \dot{F} + \dfrac{\partial \Psi}{\partial Z}\cdot \dot{Z},
\end{equation}
which, by substituting in equation (\ref{eq:thermo3}), leads to
\begin{equation}
\left( \dfrac{\partial\Psi}{\partial \Theta}+ H^{\dagger}\right) \,\dot{\Theta} +
\left( \dfrac{\partial \Psi}{\partial F} - S\right): \dot{F} - \left(\dfrac{\partial \Psi}{\partial Z}\cdot \dot{Z} +  D \right)=0.
\label{eq:equatingu}
\end{equation}
Since inequality (\ref{eq:equatingu}) must hold for every choice of the rate of state variables, the arbitrariness of $\dot{\Theta}$, $\dot{F}$, $\dot{Z}$, and volume $\mathcal{V}$ leads to the following a priori volume averaged constitutive restrictions
\begin{subequations}
	\begin{align}
	\label{eq:rel1}
	&H^{\dagger} = -\dfrac{\partial  \Psi}{\partial \Theta},\\
	\label{eq:rel2}
	&S =\dfrac{\partial  \Psi}{\partial F},\\
	\label{eq:rel3}
	&D =- T_k \cdot \dot{Z}_k, \quad \text{ with } \quad T_k=\frac{\partial\Psi}{\partial Z_k}
	\end{align}
\end{subequations}
where $T_k$ are the thermodynamic stresses, conjugate to $Z_k$. It is worth noticing that the aforementioned framework can be easily extended to generalized continuum theories \citep{germain1973method,forest2020continuum}.\\

\indent For the sake of generality we do not specify neither the physical nor the tensorial nature of ISV. However, we recall that internal state variables cannot be chosen arbitrarily, but rather need to fulfill some requirements \citep[see][]{muschik}. In particular, ISV and their rates do not occur in the balance equation of the internal energy, i.e., Eq. (\ref{eq:thermo0}). 
Furthermore, internal state variables require evolution laws, constrained by the second law of thermodynamics. Note that, for non-dissipative processes, the rates and the values of internal state variables are not necessarily zero (for $N_{Z}>1$), but such that expression (\ref{eq:rel3}) satisfies $D=0$.\\
\indent In the following, we shall consider as candidate for the description by an internal variable, any feature of the microstructure, satisfying the aforementioned requirements and undergoing irreversible rearrangements during changes of the state variables $\chi$ \citep{maugin1994thermodynamicsA,maugin1994thermodynamicsB}, e.g. local dislocations, voids, cracks densities, particle size or shape distributions, etc.

\noindent For simplicity, we will consider isothermal processes only, i.e., $\dot{\Theta}=0$. Nevertheless, we emphasize that the proposed approach can be applied for non-isothermal transformations as well.

\section{Learning the constitutive behavior of materials with Thermodynamics-based Artificial Neural Networks}
\label{sec:tann}
\noindent Let us consider a feed-forward artificial neural network of depth $K$ corresponding to a network with an input layer, $K-1$ hidden layers and an output layer ($K$-th layer). The signal flows from layer $(k - 1)$ to layer $(k)$ according to
\begin{equation}
\alpha^{(k)}_j = \mathcal{A}^{(k)}\left(  w^{(k)}_{jl} \alpha^{(k-1)}_l  + b^{(k)}_j\right),
\end{equation}
where $\alpha^{(k)}_j$ is the output of node $j$, at layer $(k)$; $\mathcal{A}^{(k)}$ is the activation function of layer $(k)$; $w^{(k)}_{jl}$ is the weight between the $l$-th node in layer $(k-1)$ and the $j$-th node in layer $(k)$; and $b^{(k)}_j$ is the bias of the $j$-th node in layer $(k)$. The activation function at the output layer, $\mathcal{A}^{(K)}$, is an identity function.

Weights and biases of interconnections are adjusted, in an iterative procedure \cite[gradient descent algorithm ][]{geron2019hands}, to minimize the error between benchmarks, $\bar{o}$, and predictions, $o$, that is measured by a loss function, $\ell$, and a training loss $\mathcal{L}$. In the following, the Mean (over a set of $N$ samples) Absolute Error (MAE) is used as loss function, i.e.,
\begin{equation}
\mathcal{L}(o) = \dfrac{1}{N} \sum_{i=1}^{N} \ell(o_i)  \quad \text{ with }\quad \ell(o_i) = ||\,\bar{o}_i - o_i||_1,
\label{eq:loss}
\end{equation}
where $||\cdot||_1$ denotes the L1 norm and $i=1,2,\ldots N$. The errors related to each node of the output layer are hence back-propagated to the nodes in the hidden layers and used to calculate the gradient of the loss function. This latter is used to update the weights and biases and minimize the loss function values.\\
\indent Once the ANN is trained, it is used in recall mode (inference phase) for output predictions.

\subsection{Thermodynamics-based Artificial Neural Networks with a priori determined internal state variables}
\noindent Relying on the theoretical background presented in Section \ref{sec:theory}, the authors have previously developed the so-called Thermodynamics-based Artificial Neural Networks \citep[TANN, see][]{masi2021thermodynamics}. In the following, we briefly summarize this method, which will be extended, in the next paragraph, to materials for which the internal variables are not a priori known.

\begin{figure}[h]
	\centering
	\includegraphics[width=0.8\textwidth]{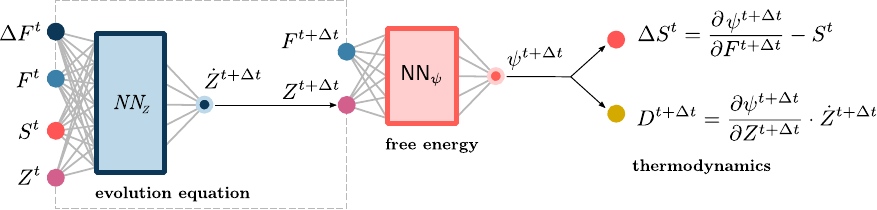}
	\caption{Structure of TANN with a priori determined internal state variables.}
	\label{fig:oldTANN}
\end{figure}

TANN rely on an incremental formulation of the material behavior. Figure \ref{fig:oldTANN} shows the architecture of the network, for isothermal material processes. The model inputs are the previous material state at time $t$, i.e., $S^t$, $F^t$ and $Z^t$, the strain increment $\Delta F^t$, and the time increment, $\Delta t$. TANN are trained to output the updated internal state variables at $t+\Delta t$, $Z^{t+\Delta t}$, the energy potential, $\Psi^{t+\Delta t}$, the stress increment, $\Delta S^t$, and the dissipation rate, $D^{t+\Delta t}$. The stress and dissipation rate are computed by respecting Equations (\ref{eq:rel2}-\ref{eq:rel3}), see \citet{masi2021thermodynamics}. The loss function $\ell$ is composed of four terms, which represent the agreement between the network outputs and the training data
\begin{equation}
\ell = \lambda^{\dot{Z}} \ell({\dot{Z}}) +\lambda^{\Psi} \ell({\Psi}) +\lambda^{\Delta S} \ell({\Delta S}) +\lambda^D \ell({D}),
\label{eq:lossTANN}
\end{equation}
where $\lambda^{\dot{Z}},\lambda^{\Psi},\lambda^{\Delta S},\lambda^D$ are weights accounting for the relative influence of each output in the total loss function, and $\ell({\dot{Z}}),\ell({\Psi}), \ell({\Delta S}), \ell({D})$ are the loss functions, defined as in (\ref{eq:loss}), of $\dot{Z}^{t+\Delta t}, \Psi^{t+\Delta t}, \Delta S^t, D^{t+\Delta t}$, respectively.


\noindent In recall mode (inference), the outputs of the model are slightly modified to deliver a material description class in Finite Element codes. In particular, TANN output the updated ISV and the updated stresses.\\
The efficiency and robustness of TANN for modeling inelastic materials was demonstrated in \cite{masi2021thermodynamics}.

\subsection{Internal coordinates and dimensionality reduction}

\noindent The existing aforementioned formulation of TANN requires the a priori knowledge of the nature and numerical values of the ISV. However, as argued by \citet{kirchner2005unifying}, ``\textit{it is neither a priori known which specific features of the microstructure characterize [such] a macroscopic internal variable, nor whether the macroscopic behavior is described sufficiently accurately by such a quantity}''. When dealing with microstructured materials, the identification of a sufficient set of internal state variables is not a trivial task and it is usually carried out heuristically by the user/modeler.\\ 
\indent Thus, the aforementioned formalism of TANN may fail in accurately capturing the constitutive response of microstructured materials if a non-adequate choice of the internal state variables is made. A remedy to this problem is presented below by the introduction of what we define herein as \textit{internal coordinates}.

\begin{figure}[h]
	\centering
	\includegraphics[width=0.35\textwidth]{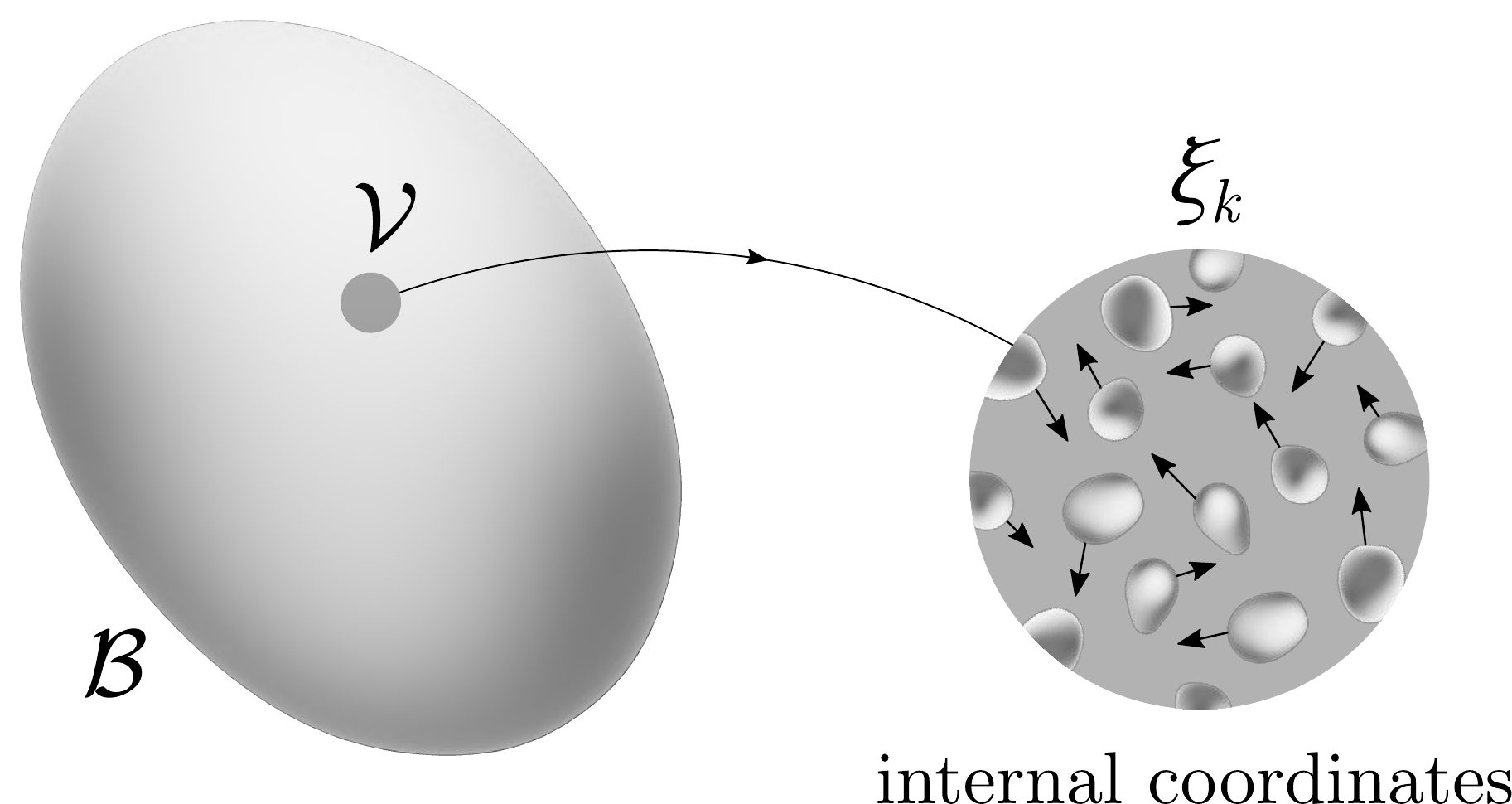}
	\caption{Schematic representation of the internal coordinates of a microstructured material.}
	\label{fig:micro}
\end{figure}

Consider a microstructured material as shown in Figure \ref{fig:micro}. We define as \textit{internal coordinates} $\xi=\{\xi_i\}$ ($i=1,\dots,N_{IC}$) the set of $N_{IC}$ microscopic quantities describing the microscopic material state. Displacement, velocity, and/or momentum fields may be regarded as internal coordinates (see Fig. \ref{fig:micro}). Other examples of internal coordinates (IC) may be any microstructural rearrangement and/or characteristic spatial and time lengths. In general, IC can be identified as microscopic quantities describing the material state \citep[a similar idea, applied in chemistry, can be also found in][]{prigogine1955thermodynamics}. Note that the nature of the IC is essentially different from that of internal state variables, which are macroscopic quantities. Moreover, internal coordinates do not necessarily coincide with the internal degrees of freedom of the assumed equivalent continuum (e.g. translational degrees of freedom in Cauchy continua, curvatures in Cosserat media).

In microstructured materials, internal coordinates usually span high-dimensional spaces, whose treatment is computationally demanding. One can thus resort to dimensionality reduction techniques \citep[see][]{kunisch2002galerkin,brunton2022data,daniel2022physics} and find a low-dimensional parametrization of the original space. 
Here, we consider autoencoders, which have been shown to be effective for studying a wide variety of physical phenomena and subjects \citep[see e.g.][]{erichson2019physics,li2020guided,fukami2020convolutional,lu2020extracting,hernandez2021deep}.
In their simplest form, autoencoders are unsupervised learning algorithms that map inputs to \textit{latent} representations of minimum dimensionality and then back to themselves. Given an input $\mathcal{I}\in \mathbb{R}^n$, we want to learn a latent representation $\mathcal{R}\in \mathbb{R}^l$ (where $l\ll n$), which is mapped back into $\mathcal{I}_*\in \mathbb{R}^n$, imposing $\mathcal{I}=\mathcal{I}_*$. The parametrization  is  implemented  by  two  functions:  an  encoder $ENC:\mathbb{R}^n\rightarrow \mathbb{R}^l$ and a decoder $DEC:\mathbb{R}^l\rightarrow \mathbb{R}^n$. The functions $ENC$ and $DEC$ are determined/learnt so to minimize the error $||\mathcal{I}-\mathcal{I}_*||$ (reconstruction error). As $l\ll n$, dimensionality reduction is achieved.

Dimensionality reduction algorithms can thus be used to find a lower-dimensional representation of the internal coordinates of complex, microstructured materials. However, the latent representation obtained by minimization of the representation error (in the case of autoencoders, for instance) does not necessarily respect the physics of the problem. Indeed, dimensionality reduction methods are not thermodynamics-aware, hence do not assure that the thermodynamic structure of the problem, as detailed in subsection \ref{subsec:state}, is preserved. A remedy to this consists of enforcing the parametrization of the internal coordinates within the formalism of TANN.

\subsection{Thermodynamics-based Artificial Neural Networks without a priori determined internal state variables}
\label{subsec:TANNISV}
\noindent We propose an extension of the previous developed Thermodynamics-based Artificial Neural Networks in order to account for the evolution of internal coordinates of materials with microstructure. This is accomplished using internal state variables. The internal state variables are identified automatically via latent representations of the internal coordinates. Therefore, the extended TANN do not require any a priori choice of the nature and values of the internal state variables. In this sense, learning is unsupervised.\\
\begin{figure}[t]
	\centering
	\includegraphics[width=\textwidth]{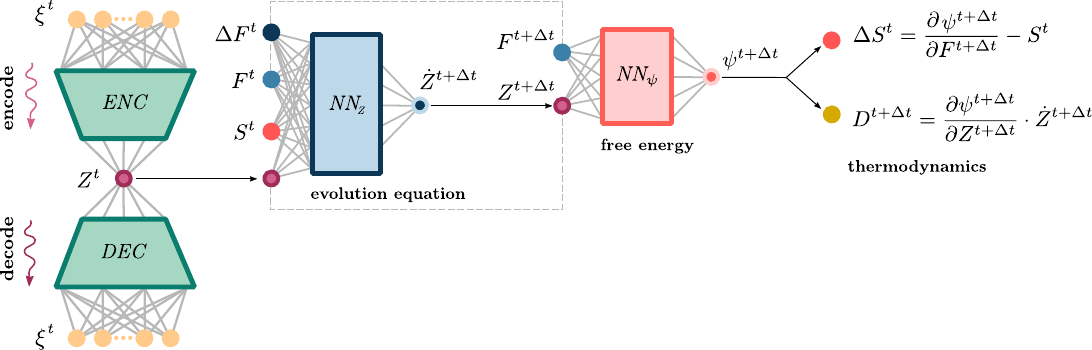}
	\caption{Structure of TANN without a priori determined internal state variables.}
	\label{fig:TANNnew}
\end{figure}
Figure \ref{fig:TANNnew} shows the architecture of the extended TANN. The training and recall algorithms are sketched below in Algorithms \ref{alg:2} and \ref{alg:3}. By comparing the extended scheme with the previous one (Fig. \ref{fig:oldTANN}), we see that the difference lies on the introduction of an encoder $ENC$ and a decoder $DEC$. The encoder and decoder reduce the high dimensionality of the internal coordinates to a minimum number of variables (latent variables), which we identify later as the ISV of the microstructure.

In particular, the encoder is fed with the coordinates $\xi^t$ and outputs a set of latent variables, at time $t$. The latter are then used to feed the original algorithm of TANN. The decoder performs the inverse parametrization and outputs the updated internal coordinates, $\xi^{t+\Delta t}$, from the knowledge of the updated latent variables. The identified variables are arguments of the free-energy and of the dissipation rate. As such, they fulfill the thermodynamic restrictions of the first and second law of thermodynamics. These properties make the latent variables to be a thermodynamically admissible set of internal state variables.

\begin{center}
\begin{algorithm}[H]
\scriptsize

\DontPrintSemicolon
  \SetKwFunction{FMain}{}
  \SetKwProg{Fn}{TANN}{:}{}
  \Fn{\FMain{$\mathcal{I}$}}{
        \vspace{0.05cm} $\mathcal{I}=\{F^t, S^t, \xi^t, \Delta F^t, \Delta t\}$\;
        \vspace{0.1cm}
        
\begin{minipage}{.9\linewidth}
         \DontPrintSemicolon
  \SetKwFunction{FMain}{}
  \SetKwProg{Fn}{$\blacktriangleright$ Encoder $ENC(\mathcal{I}^t_{ENC})$ :}{}{}
  \Fn{}{\vspace{0.05cm}
         $\mathcal{I}^t_{ENC} = \{\xi^t\}$\;
        \KwRet $Z^t$\;
  }
  \end{minipage}\vspace{0.1cm}             

\begin{minipage}{.7\linewidth}
\DontPrintSemicolon
  \SetKwFunction{FMain}{}
  \SetKwProg{Fn}{$\blacktriangleright$ Neural Network $NN_Z\left(\mathcal{I}_Z\right)$ :}{}{}
  \Fn{}{\vspace{0.05cm}
         $\mathcal{I}_Z = \{F^t, S^t, Z^t, \Delta F^t, \Delta t\}$\;
        \KwRet $\dot{Z}^t$\;
  }
  \end{minipage}   
  
 $\vartriangleright$ updated internal state variables (finite difference): $Z^{t+\Delta t} = Z^t + \dot{Z}^t\Delta t$\vspace{0.1cm}

$\vartriangleright$ updated state variables (definition): $F^{t+\Delta t} = F^t + \Delta F^t$\vspace{0.1cm}

\begin{minipage}{.7\linewidth} 
  \DontPrintSemicolon
  \SetKwFunction{FMain}{}
  \SetKwProg{Fn}{$\blacktriangleright$ Neural Network $NN_{\Psi}$ :}{}{}
  \Fn{\FMain{\vspace{0.05cm} $\mathcal{I}_{\Psi}$}}{
         $\mathcal{I}_{\Psi} = \{F^{t+\Delta t},Z^{t+\Delta t}\}$\;
        \KwRet $\Psi^{t+\Delta t}$\;
  }
  \end{minipage}
  
\begin{minipage}{.7\linewidth}
\DontPrintSemicolon
  \SetKwFunction{FMain}{}
  \SetKwProg{Fn}{$\blacktriangleright$ Thermodynamic restrictions :}{}{}
  \Fn{}{
         stress increment, Eq. (\ref{eq:rel2}): $\Delta S^t = \frac{\partial \Psi^{t+\Delta t}}{\partial F^{t+\Delta t}}-S^t$\;
        dissipation rate, Eq. (\ref{eq:rel3}): $D^{t+\Delta t} = - \frac{\partial\Psi^{t+\Delta t}}{\partial Z^{t+\Delta t}} \cdot \dot{Z}^{t+\Delta t}$\;
        \KwRet $\Delta S^t,D^{t+\Delta t} $\;
  }
\end{minipage}
  
\begin{minipage}{.9\linewidth}
  \DontPrintSemicolon
  \SetKwFunction{FMain}{}
  \SetKwProg{Fn}{$\blacktriangleright$ Decoder $DEC(\mathcal{I}_{DEC})$ :}{}{}
  \Fn{}{\vspace{0.05cm}
         $\mathcal{I}_{DEC} = \{Z^{t},Z^{t+\Delta t}\}$\;
        \KwRet $\xi_*^{t},\xi_*^{t+\Delta t}$\;
  }

$\vartriangleright$ compute $\nabla^t_{ED}= \frac{\partial \xi_{*i}^{t}}{\partial \xi^{t}}$\;
\end{minipage}\vspace{0.1cm}

        \KwRet $\mathcal{O}=\{\Psi^{t+\Delta t},\Delta S^t, D^{t+\Delta t}\}$ and $\mathcal{O}_{IC}=\{\xi_*^{t},\nabla^t_{ED},\xi_*^{t+\Delta t}\}$\;
  }

\caption{TANN without a priori determined internal state variables, first-step training.}
\label{alg:2}
\end{algorithm}
\end{center}

\noindent The training of the extended TANN involves a two-step procedure.
\subparagraph{Training.} The first step, sketched in Algorithm \ref{alg:2}, allows to find and compute the internal state variables associated with the evolution of the internal coordinates of a material. The loss function is written as
\begin{equation}
\ell = \lambda^{\xi} \left[\ell({\xi}_*^t)+\ell({\xi}_*^{t+\Delta t})\right]+\lambda^{\nabla^t_{ED}} \ell(\nabla^t_{ED}) +\lambda^{\Psi} \ell({\Psi}) +\lambda^{\Delta S} \ell({\Delta S}) +\lambda^D \ell({D}),
\end{equation}
with $\lambda^{\xi},\lambda^{\nabla^t_{ED}}$ being the weights accounting for the relative influence of outputs of the encoder and decoder; $\ell({\xi}_*^t)$ and  $\ell({\xi}_*^{t+\Delta t})$ the loss functions of the reconstructed internal coordinates; and $\ell(\nabla^t_{ED})$ is the loss function of the differentiation of the decoder's outputs with respect to the encoder's inputs, i.e., $\nabla^t_{ED}= \partial \xi_{*i}^{t}/\partial \xi^{t}$, which, by definition, must equal the identity matrix.\\
Differently from expression (\ref{eq:lossTANN}), there is no loss associated to the prediction of the internal variable rates. Indeed, no information, at this first stage of the training, is available about the internal state variables and their values. Furthermore, the decoder is constructed to perform the exact inverse transformation of the encoder, by imposing its weights to be the transpose of the encoder's weights \citep[see][]{geron2019hands}. This allows to reduce the number of learnable parameters to almost the half. It should be noted, however, that such a procedure is not necessary for obtaining small reconstruction errors.
\subparagraph{Fine tuning.} The second training step involves the fine-tuning of the neural networks $NN_Z$ and $NN_{\Psi}$, on the basis of the discovered ISV. In particular, weights and biases of the decoder are freezed, while the trained encoder is used once, before the fine-tuning, to generate the internal state variables, which become the inputs of TANN, see Algorithm \ref{alg:3} ($tuning=True$). Both during fine-tuning and recall mode, the encoder is no more used to make predictions and is detached from the rest of the network. The training thus involves the loss function in (\ref{eq:lossTANN}).

\begin{center} 
\begin{algorithm}[H]
\scriptsize
\DontPrintSemicolon
  \SetKwFunction{FMain}{}
  \SetKwProg{Fn}{TANN}{:}{}
  \Fn{\FMain{$\mathcal{I}$}}{
        \vspace{0.05cm} $\mathcal{I}=\{F^t, S^t, Z^t, \Delta F^t, \Delta t\}$\;
        \vspace{0.1cm}          

\begin{minipage}{.7\linewidth}
\DontPrintSemicolon
  \SetKwFunction{FMain}{}
  \SetKwProg{Fn}{$\blacktriangleright$ Neural Network $NN_Z\left(\mathcal{I}_Z\right)$ :}{}{}
  \Fn{}{\vspace{0.05cm}
         $\mathcal{I}_Z = \{F^t, S^t, Z^t, \Delta F^t, \Delta t\}$\;
        \KwRet $\dot{Z}^t$\;
  }
  \end{minipage}   
  
 $\vartriangleright$ updated internal state variables (finite difference): $Z^{t+\Delta t} = Z^t + \dot{Z}^t\Delta t$\vspace{0.1cm}

$\vartriangleright$ updated state variables (definition): $F^{t+\Delta t} = F^t + \Delta F^t$\vspace{0.1cm}

\begin{minipage}{.7\linewidth} 
  \DontPrintSemicolon
  \SetKwFunction{FMain}{}
  \SetKwProg{Fn}{$\blacktriangleright$ Neural Network $NN_{\Psi}$ :}{}{}
  \Fn{\FMain{\vspace{0.05cm} $\mathcal{I}_{\Psi}$}}{
         $\mathcal{I}_{\Psi} = \{F^{t+\Delta t},Z^{t+\Delta t}\}$\;
        \KwRet $\Psi^{t+\Delta t}$\;
  }
  \end{minipage}
  
\begin{minipage}{.7\linewidth}
\DontPrintSemicolon
  \SetKwFunction{FMain}{}
  \SetKwProg{Fn}{$\blacktriangleright$ Thermodynamic restrictions :}{}{}
  \Fn{}{
         stress increment, Eq. (\ref{eq:rel2}): $\Delta S^t = \frac{\partial \Psi^{t+\Delta t}}{\partial F^{t+\Delta t}}-S^t$\;
        dissipation rate, Eq. (\ref{eq:rel3}): $D^{t+\Delta t} = - \frac{\partial\Psi^{t+\Delta t}}{\partial Z^{t+\Delta t}} \cdot \dot{Z}_k^{t+\Delta t}$\;
        \KwRet $\Delta S^t,D^{t+\Delta t} $\;
  }
\end{minipage}
  
\begin{minipage}{.9\linewidth}
  \DontPrintSemicolon
  \SetKwFunction{FMain}{}
  \SetKwProg{Fn}{$\blacktriangleright$ Decoder $DEC(\mathcal{I}_{DEC})$ :}{}{}
  \Fn{}{\vspace{0.05cm}
         $\mathcal{I}_{DEC} = \{Z^{t+\Delta t}\}$\;
        \KwRet $\xi_*^{t+\Delta t}$\;
  }
\end{minipage}\vspace{0.1cm}
 
    \uIf{tuning=True}{
    $\mathcal{O}=\{\dot{Z}^t, \Psi^{t+\Delta t},\Delta S^t, D^{t+\Delta t}\}$\;
  }
  \Else{
    $\mathcal{O}=\{F^{t+\Delta t}, Z^{t+\Delta t}, S^{t+\Delta t},\Psi^{t+\Delta t}, D^{t+\Delta t}, \xi_*^{t+\Delta t}\}$ \;
  }
        \KwRet $\mathcal{O}$\;   

  }

\caption{TANN without a priori determined internal state variables.}
\label{alg:3}
\end{algorithm}
\end{center}

Once trained, predictions are made using Algorithm \ref{alg:3} ($tuning=False$, recall mode). This algorithm makes use of the decoder in recall mode in order to predict the IC based on the internal state variables at any material state. Notice, that if the reconstruction  of the internal coordinates is not needed, then the decoder can be omitted in recall mode. This could be the case in large multiscale analyses, where the microstructural fields are not of interest. In this case the trained network operates in recall mode as the classical TANN.

It is worth emphasizing that the process histories of the evolution of the system are not needed in this formulation. This is an important advantage compared to other data-driven methods as far it concerns the data generation and the efficiency of large multiscale simulations. 

\section{Applications to microstructured, lattice materials}
\label{sec:app1}
\noindent In the following, we shall consider, as benchmark, lattice cells, with inelastic microstructure, undergoing small strains, namely $|\nabla u|\ll 1$, $p\approx \sigma$, and $\text{Sym}(f)\approx \varepsilon-I$, $\varepsilon=\text{Sym}(\nabla u)$, with $\sigma$ being the Cauchy stress tensor, $\varepsilon$ the infinitesimal strain tensor, and $\text{Sym}(f)$ denoting the symmetric part of the deformation gradient.

The microscopic incremental problem on the lattice cell, $\mathcal{V}$, is solved in terms of the microscopic displacement field, by a code developed by the authors (see Appendix A). For any microscopic quantity, volume averages are computed with respect to the volume $\mathcal{V}$.\\
The average stress tensor is equal to:
\begin{equation}
\Sigma = \langle\,\sigma \rangle = \frac{1}{|\mathcal{V}|} \int_{\mathcal{V}} \sigma \, dx = \frac{1}{|\mathcal{V}|} \sum_i^{N_{nodes}^p} t_i \otimes  x_{\Delta i}^p,
\end{equation}
where $\sigma$ is the microscopic stress tensor; $t$ the surface traction vectors; $N_{nodes}^p$ the boundary nodes; and $x_{\Delta}^p=x^{p+}-x^{p-}$. The superscripts $p+$ and $p-$ denote values evaluated at opposite boundaries. In smalls strains, $S \approx \Sigma$.\\
The average strain tensor is given by 
\begin{equation}
E = \langle\,\varepsilon \rangle = \frac{1}{|\mathcal{V}|} \int_{\mathcal{V}} \varepsilon \, dx = \frac{1}{|\mathcal{V}|} \int_{\partial \mathcal{V}} u \otimes x\, ds,
\end{equation}
where $\varepsilon$ is the microscropic strain tensor and $x$ and $u$ are the position and the displacement field at the boundaries, respectively.\\
\indent Periodic boundary conditions are considered in order to use our approach for the BVP in the next Section, using asymptotic homogenization, incrementally.

\begin{figure}[h]
	\centering
	\includegraphics[width=0.5\textwidth]{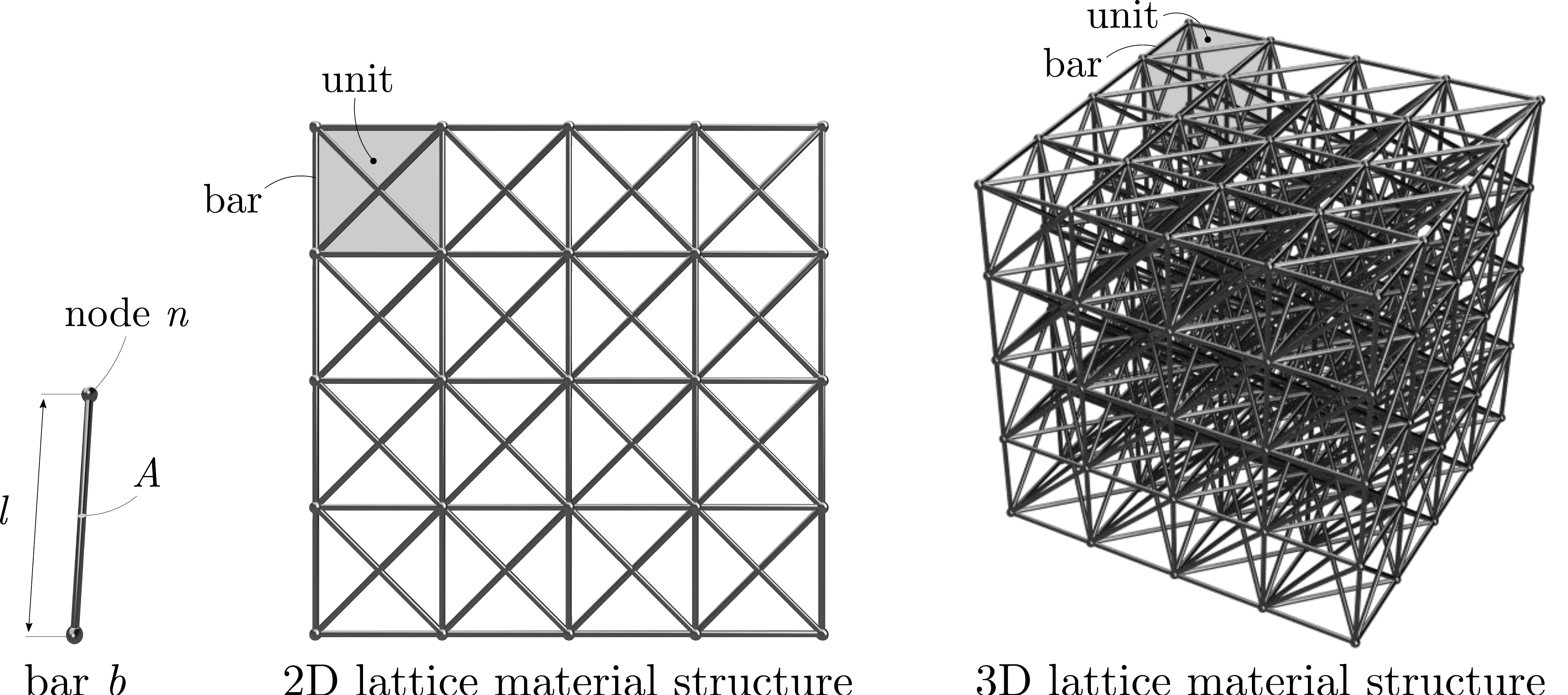}
	\caption{Examples of 2D and 3D lattice material structures.}
	\label{fig:lattice1}
\end{figure}

In the applications presented in this work, we consider two-(2D) and three-dimensional (3D) lattice material structures made of 6- and 40-elements (bars), respectively (see Fig. \ref{fig:lattice1}). The bars display an elasto-plastic behavior, with strain-hardening \citep[see][]{masiTANNspigl}, with Young modulus $100$ GPa, yield strength $300$ MPa, and hardening parameter $10$ GPa. Each bar has a circular cross-section $A=1$ cm$^2$, see Fig. \ref{fig:lattice1}.

Internal coordinates are considered to be the microscopic displacements, at each node $n$ (Fig. \ref{fig:lattice1}), and the (uniform) internal forces along each bar $b$. However, the internal coordinates $\xi$, and, consequently, our approach are not limited to this particular choice.

\subsection{Data generation}
\label{subsec:datagen}
\noindent Data are generated by identifying an initial, reference state for the material at time $t=0$, with initial zero strain and stress, and by applying random strain increments with a uniform standard distribution, assumed constant over the individual time step, $\Delta t$. The randomly generated material states are then divided into two sets of increments, one at time $t$ and the other at time $t+\Delta t$. The data-sets are randomly shuffled, as the evolution of the material state is fully described by the internal state variables and process histories are not needed.\\
Additionally, artificial subsets are constructed assuming zero strain increments from a partial set ($20\%$) of the initially generated data. Such artificial data are added to the training set with the objective of making the trained network able to associate no changes in the material state for zero strain increments.\\
The average stress, $\Sigma$, strain, $E$, total free-energy, $\Psi$, and total dissipation rate, $D$, are computed.\\
For 2D material structures, we generate a total of 30'000 data, while for the 3D case, a total of 50'000 data.

\noindent The hyper-parameters (i.e., number of hidden layers, neurons, activation functions, etc.) of the networks are selected to give the best predictions. This is accomplished by comparing the learning error on the set of test patterns, per each trial choice of the hyper-parameters. In each training process, we use early-stopping \citep{geron2019hands}. Adam optimizer with Nesterov’s acceleration gradient is selected and a batch size of 100 samples is used. All data are normalized between -1 and 1.

\subsection{2D regular lattices}
\label{subsec:2Dref}
\begin{figure}[h]
	\centering
	\includegraphics[width=0.5\textwidth]{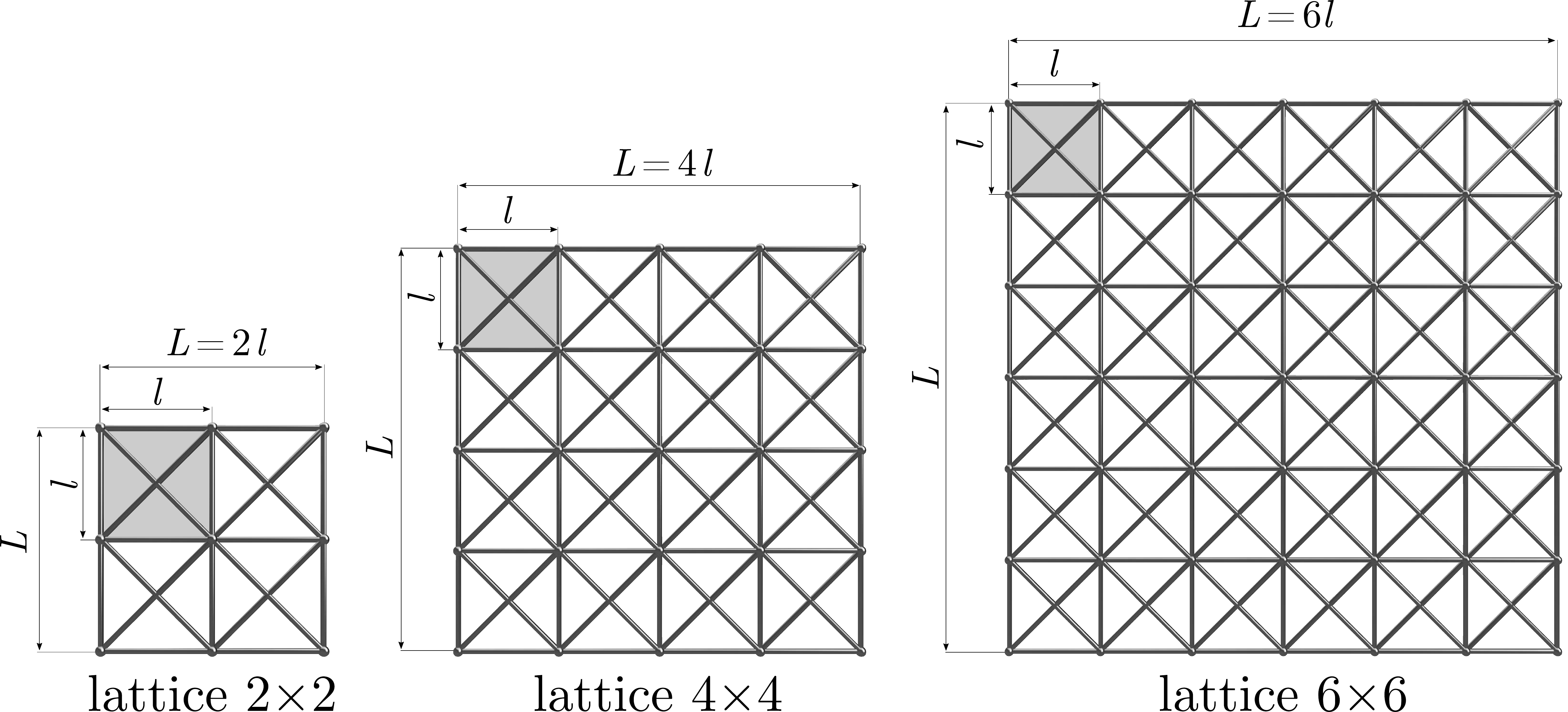}
	\caption{Investigated 2D regular lattices.}
	\label{fig:lattice2Dreg}
\end{figure}

\noindent We start by considering 2D regular lattices, of dimensions $L\times L$, with $L=1$ m, composed of $2\times 2$, $4\times 4$, and $6\times 6$ units, see Figure \ref{fig:lattice2Dreg}.

\noindent The architecture of TANN is kept constant in all cases. In particular, the encoder and decoder have depth $K=4$, with a total of 372 nodes, each one, and ELU activation functions \citep{clevert2015fast}. Network $NN_Z$ has $K=3$, each hidden layer has 56 nodes and leaky ReLU activations \citep{maas2013rectifier}. Network $NN_{\Psi}$ has depth $K=1$, 48 (hidden) nodes, and fourth order RePU activation functions \citep{li2019better}. All output layers have linear activation functions, and those of networks $NN_Z$ and $NN_{\Psi}$ have, additionally, zero biases.\\
From the encoding of the internal coordinates, we find 12 ISV, which is much less than the number of internal coordinates of the lattice structures.

\begin{figure}[h]
	\centering
\begin{subfigure}[t]{\textwidth}
  \centering
  \includegraphics[width=\linewidth]{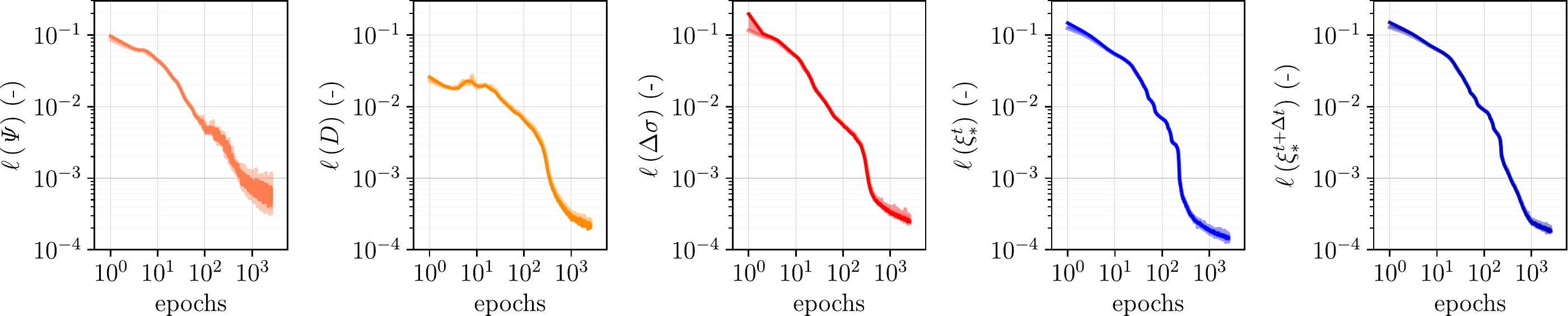}
	\caption{\footnotesize first step training}
\end{subfigure}\\ \vspace{0.3cm}
\begin{subfigure}[t]{\textwidth}
  \centering
  \includegraphics[width=0.8\linewidth]{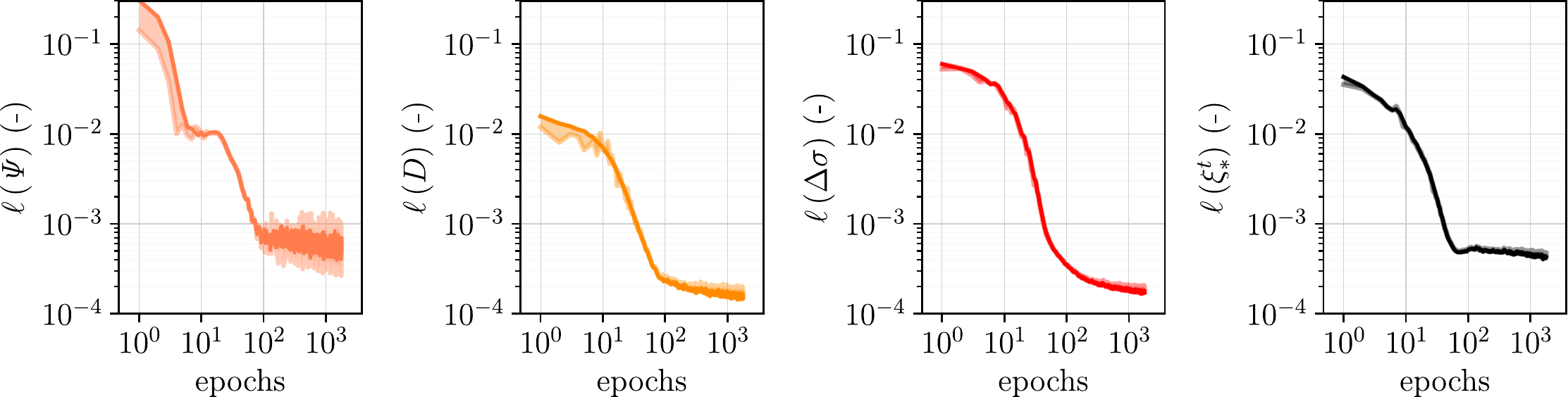}
	\caption{\footnotesize second step training, fine-tuning}
\end{subfigure}
	\caption{Dimensionless Mean Absolute Error (MAE) of TANN predictions, in function of the training epochs, evaluated with respect to the training set and the validation one (shaded regions), for: (a) first step and (b) second step training$-$fine-tuning (see paragraph \ref{subsec:TANNISV}, Algorithms \ref{alg:2}-\ref{alg:3}).}
	\label{fig:training}
\end{figure}

\noindent Figure \ref{fig:training} displays the training and validation errors in function of training epochs, within the two-step training procedure (see paragraph \ref{subsec:TANNISV}, Algorithms \ref{alg:2}-\ref{alg:3}). The early stopping rule assures convergence, after approximately 2500 and 1000 epochs, for the first and second step, respectively.

After training, TANN are used in recall mode, i.e., to make predictions for never-seen strain loading paths. As discussed in paragraph \ref{subsec:TANNISV}, the trained network is independent of the values of the internal coordinates. TANN are fed with strain, stress, and discovered internal state variables, at time $t$, and predict the material state at time $t+\Delta t$ (cf. paragraph \ref{subsec:state}). Additionally, the network, relying on the trained decoder, can output the updated values of the internal coordinates. This feature allows one to track microstructural evolutions, at extremely reduced computational cost \citep[in contrast with multiscale simulations;][]{Feyel2003,Nitka2011,Eijnden2016}.

\begin{figure*}[b]
\centering
\begin{subfigure}[t]{0.2\textwidth}
  \centering
  \includegraphics[width=\linewidth]{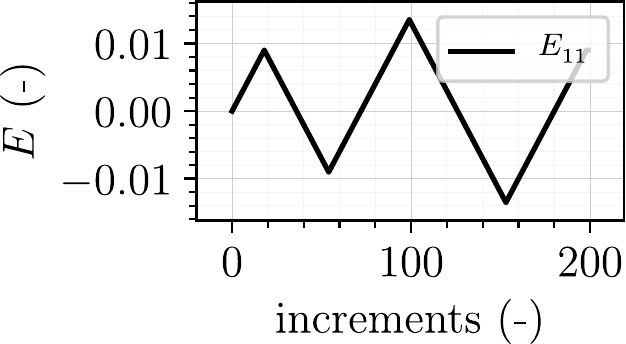}
	\caption{\footnotesize strain path}
	\label{fig:L2x2_1a}
\end{subfigure} \hspace{0.3cm}
\begin{subfigure}[t]{0.45\textwidth}
  \centering
  \includegraphics[width=\linewidth]{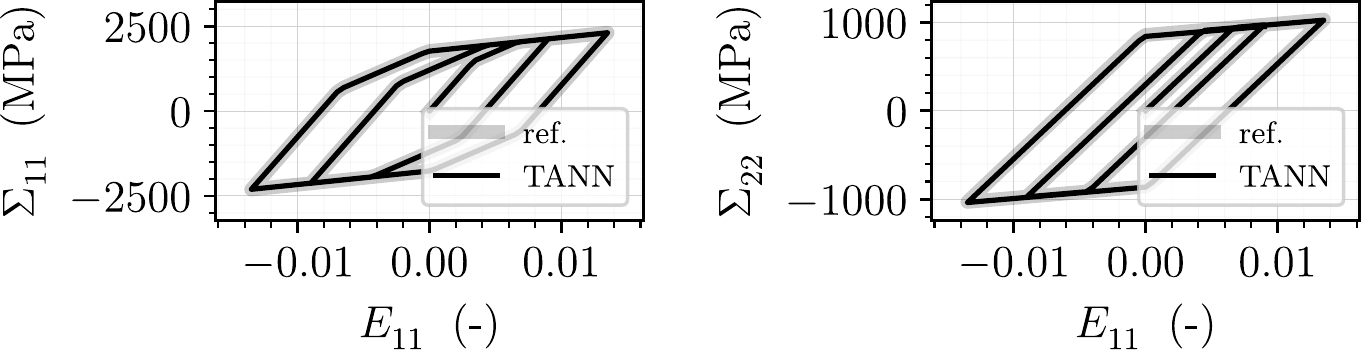}
	\caption{\footnotesize stress-strain behavior}
	\label{fig:L2x2_1b}
\end{subfigure}\\ \vspace{0.3cm}
\begin{subfigure}[t]{0.45\textwidth}
  \centering
  \includegraphics[width=\linewidth]{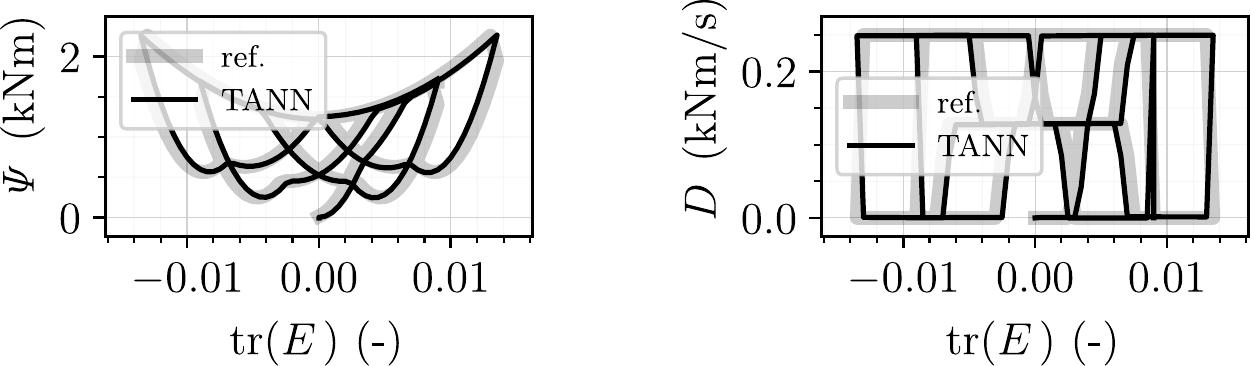}
	\caption{\footnotesize free-energy and dissipation rate in function of volume change}
	\label{fig:L2x2_1c}
\end{subfigure}\\ \vspace{0.3cm}
\begin{subfigure}[t]{0.8\textwidth}
  \centering
  \includegraphics[width=\linewidth]{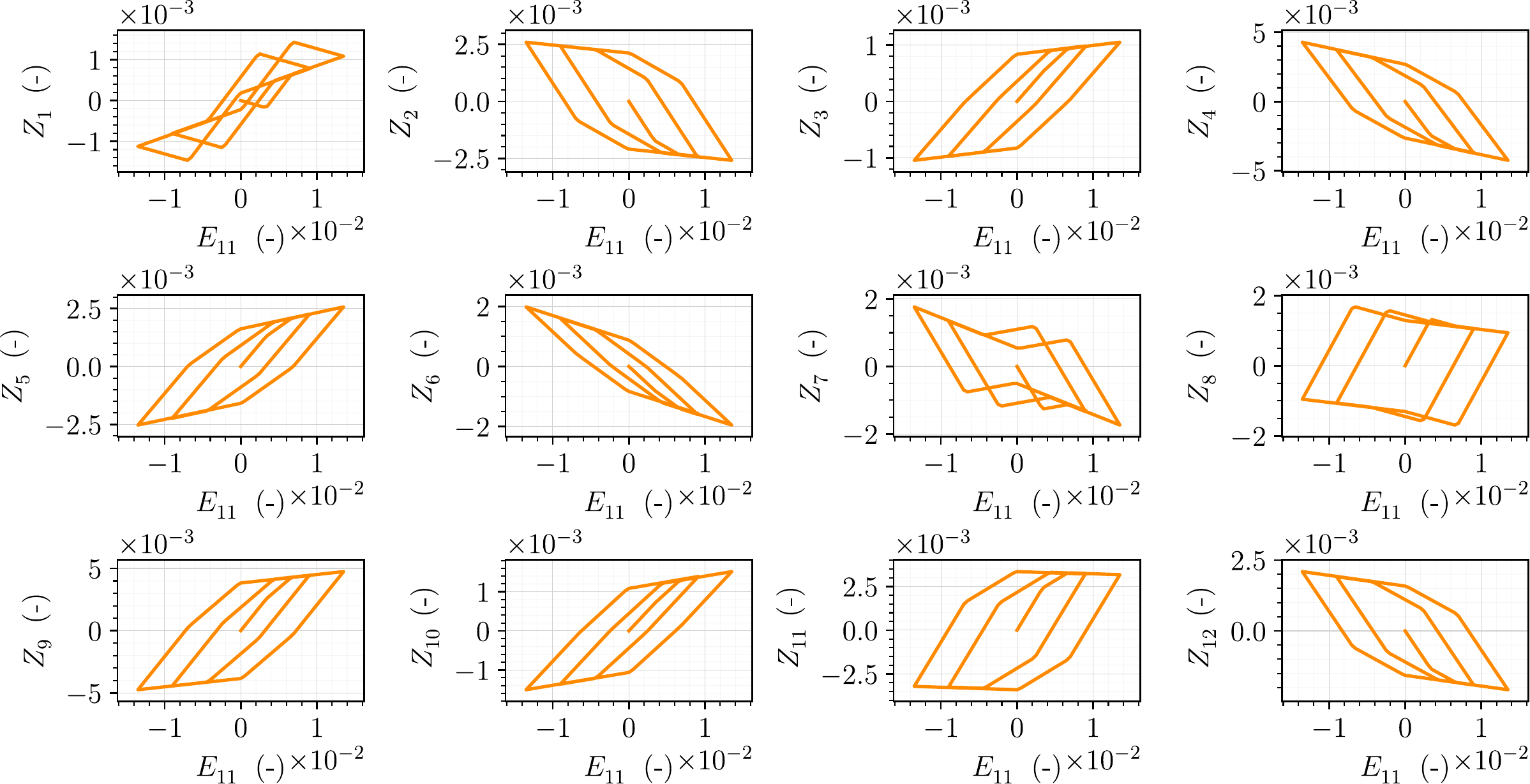}
	\caption{\footnotesize internal state variables in function of strain}
	\label{fig:L2x2_1d}
\end{subfigure}\\ \vspace{0.3cm}
\begin{subfigure}[t]{0.45\textwidth}
  \centering
  \includegraphics[width=\linewidth]{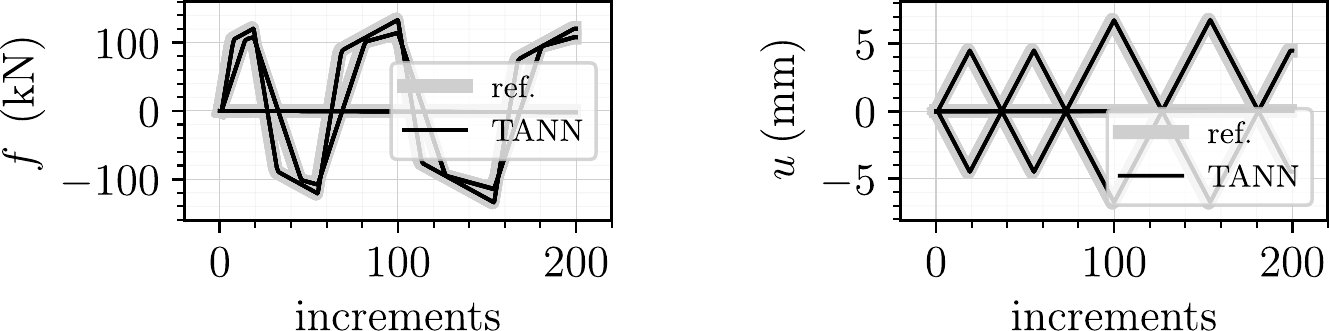}
	\caption{\footnotesize internal coordinates: internal forces $f$ (left) and microscopic displacements $u$ (right)}
	\label{fig:L2x2_1e}
\end{subfigure}
\caption{Response of the $2\times 2$ lattice in terms of stresses (b), energy and dissipation rate (c), internal state variables (d), and internal coordinates (e), for a uni-axial cyclic loading path, with increasing amplitude (a).}
\label{fig:L2x2_1}
\end{figure*}

We show in Figure \ref{fig:L2x2_1} the predictions of the material response of a $2\times 2$ lattice, in terms of volume average microscopic stress, free-energy, dissipation rate, and internal state variables, under a uni-axial strain loading path with increasing amplitude (Figure \ref{fig:L2x2_1a}). The network displays extremely good performance with respect to the reference model. As extensively discussed in \citet{masi2021thermodynamics}, the predictions of TANN are always physically consistent, respecting the first and second law of thermodynamics.\\
\indent After having predicted the average material response, TANN, and namely the decoder, are used to compute the evolution of the internal coordinates, that is, the microscopic displacements and internal forces (see Fig. \ref{fig:L2x2_1d}). We observe the accurate predictions of TANN, when compared with the microscopic reference simulations, in predicting not only average quantities, such as stresses and internal state variables, but also purely microscopic ones, i.e., internal coordinates.

\begin{figure*}[h]
\centering
\begin{subfigure}[t]{0.2\textwidth}
  \centering
  \includegraphics[width=\linewidth]{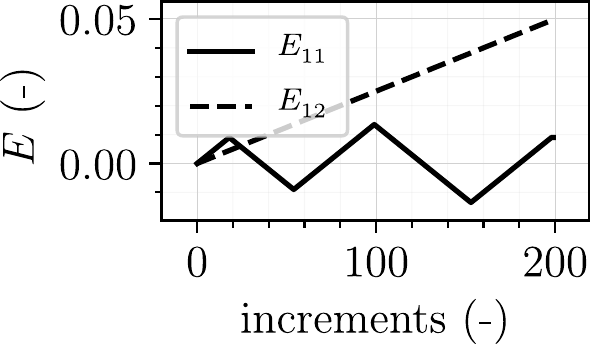}
	\caption{\footnotesize strain path}
	\label{fig:L2x2_2a}
\end{subfigure} \hspace{0.3cm}
\begin{subfigure}[t]{0.45\textwidth}
  \centering
  \includegraphics[width=\linewidth]{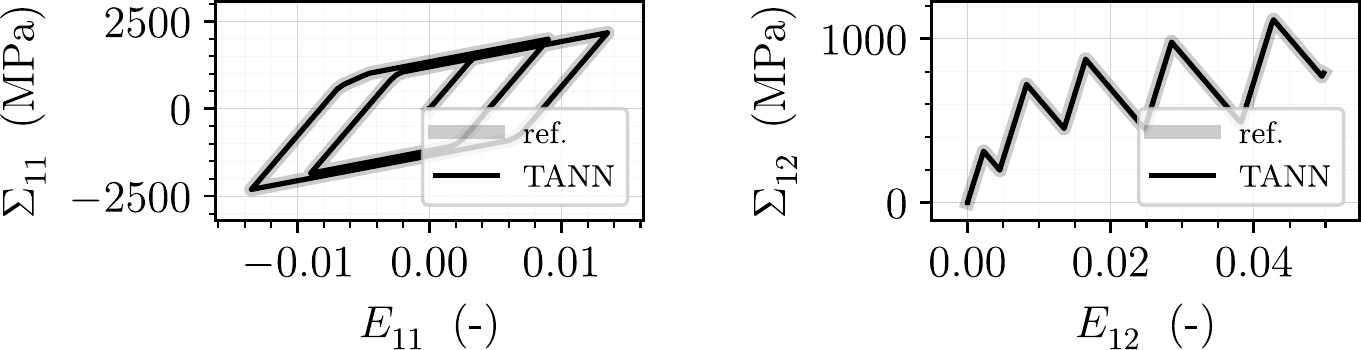}
	\caption{\footnotesize stress-strain behavior}
	\label{fig:L2x2_2b}
\end{subfigure}\\ \vspace{0.3cm}
\begin{subfigure}[t]{0.8\textwidth}
  \centering
  \includegraphics[width=\linewidth]{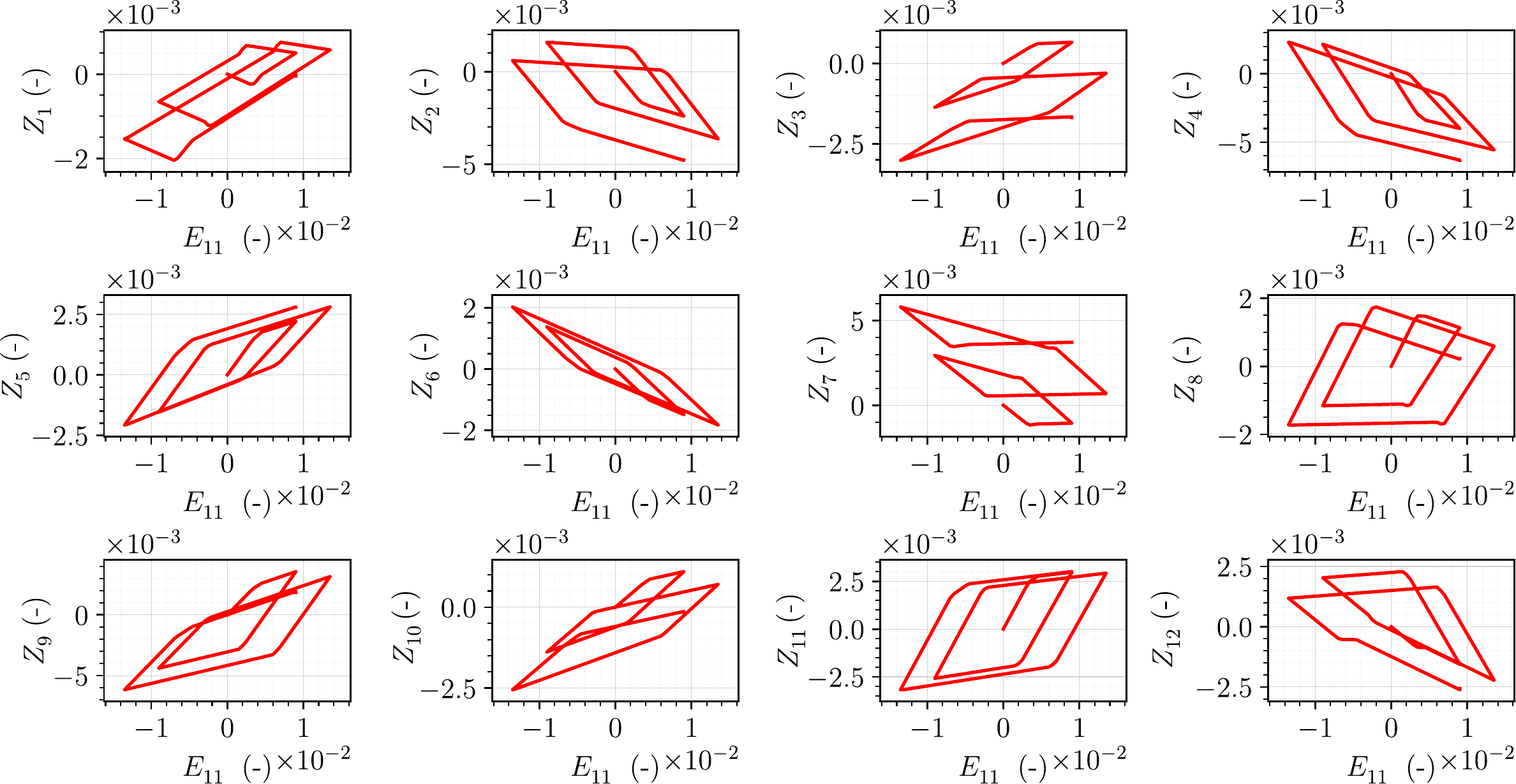}
	\caption{\footnotesize internal state variables in function of strain}
	\label{fig:L2x2_2c}
\end{subfigure}\\ \vspace{0.3cm}
\begin{subfigure}[t]{0.45\textwidth}
  \centering
  \includegraphics[width=\linewidth]{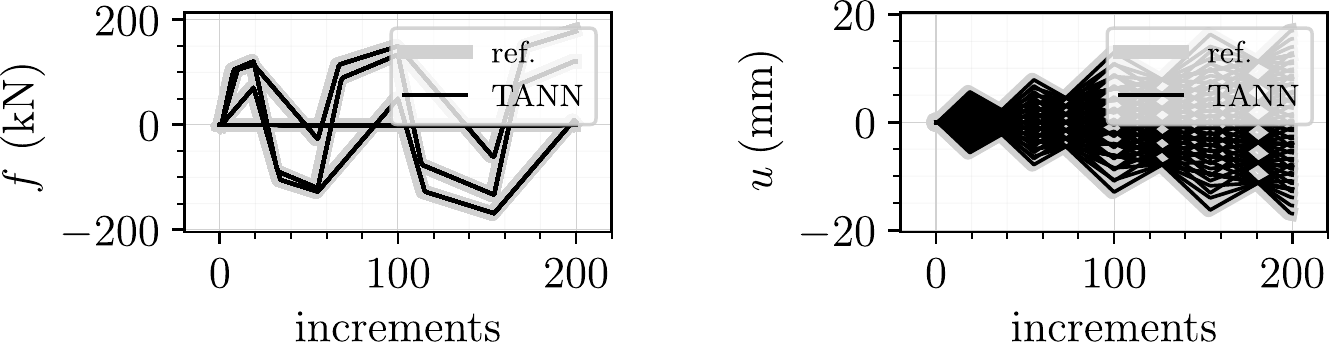}
	\caption{\footnotesize internal coordinates: internal forces $f$ (left) and microscopic displacements $u$ (right)}
	\label{fig:L2x2_2d}
\end{subfigure}
\caption{Response of the $2\times 2$ lattice in terms of stresses (b), energy and dissipation rate (c), internal state variables (d), and internal coordinates (e), for a uni-axial cyclic loading path, with increasing amplitude, and monotonous shearing (a).}
\label{fig:L2x2_2}
\end{figure*}

Figure \ref{fig:L2x2_2} displays the material predictions for a combined uni-axial strain loading (as in Figure \ref{fig:L2x2_1}) and a monotonous increasing shearing (on a $2\times 2$ lattice). The stress behavior is accurately predicted by the network, as well as the internal coordinates.\\
\indent By observing the effect of the additional shearing deformation in the computed internal state variables (Figure \ref{fig:L2x2_2d}), we notice that differently from the above case (Figure \ref{fig:L2x2_1e}), most of the internal state variables are found to oscillate between evolving amplitudes. The evolution of only three variables, namely $\{Z_5,Z_6,Z_{11}\}$, resemble those obtained under the uni-axial loading path. This is a direct consequence of the fact that the computed variables characterize local deformed regions of the microstructure.\\
\indent The physical nature of the computed internal state variables can be examined though feature extraction methods \citep[see][]{strofer2018data,lu2020extracting}. However, this exceeds the scope of the present work.

\begin{figure*}[h]
\centering
\begin{subfigure}[t]{0.2\textwidth}
  \centering
  \includegraphics[width=\linewidth]{Figure8_1.pdf}
	\caption{\footnotesize strain path}
	\label{fig:L6x6_2a}
\end{subfigure} \hspace{0.3cm}
\begin{subfigure}[t]{0.45\textwidth}
  \centering
  \includegraphics[width=\linewidth]{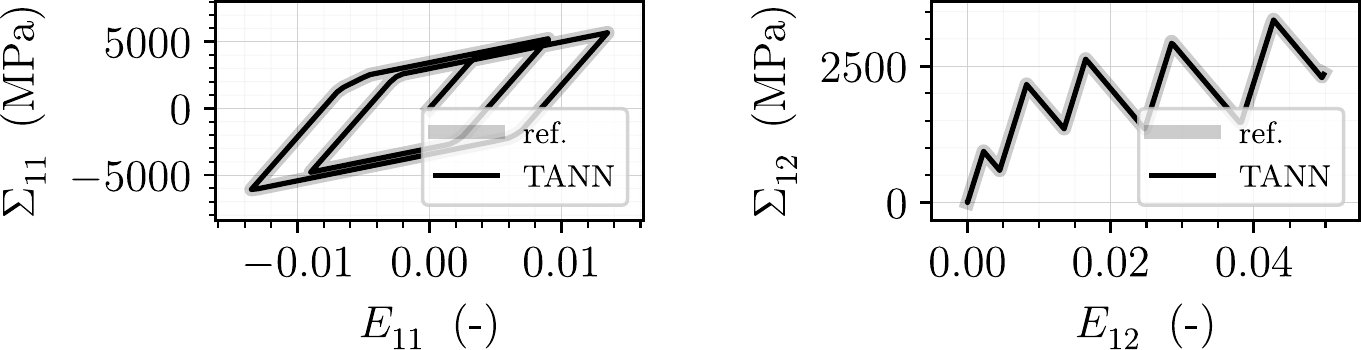}
	\caption{\footnotesize stress-strain behavior}
	\label{fig:L6x6_2b}
\end{subfigure}\\ \vspace{0.3cm}
\begin{subfigure}[t]{0.8\textwidth}
  \centering
  \includegraphics[width=\linewidth]{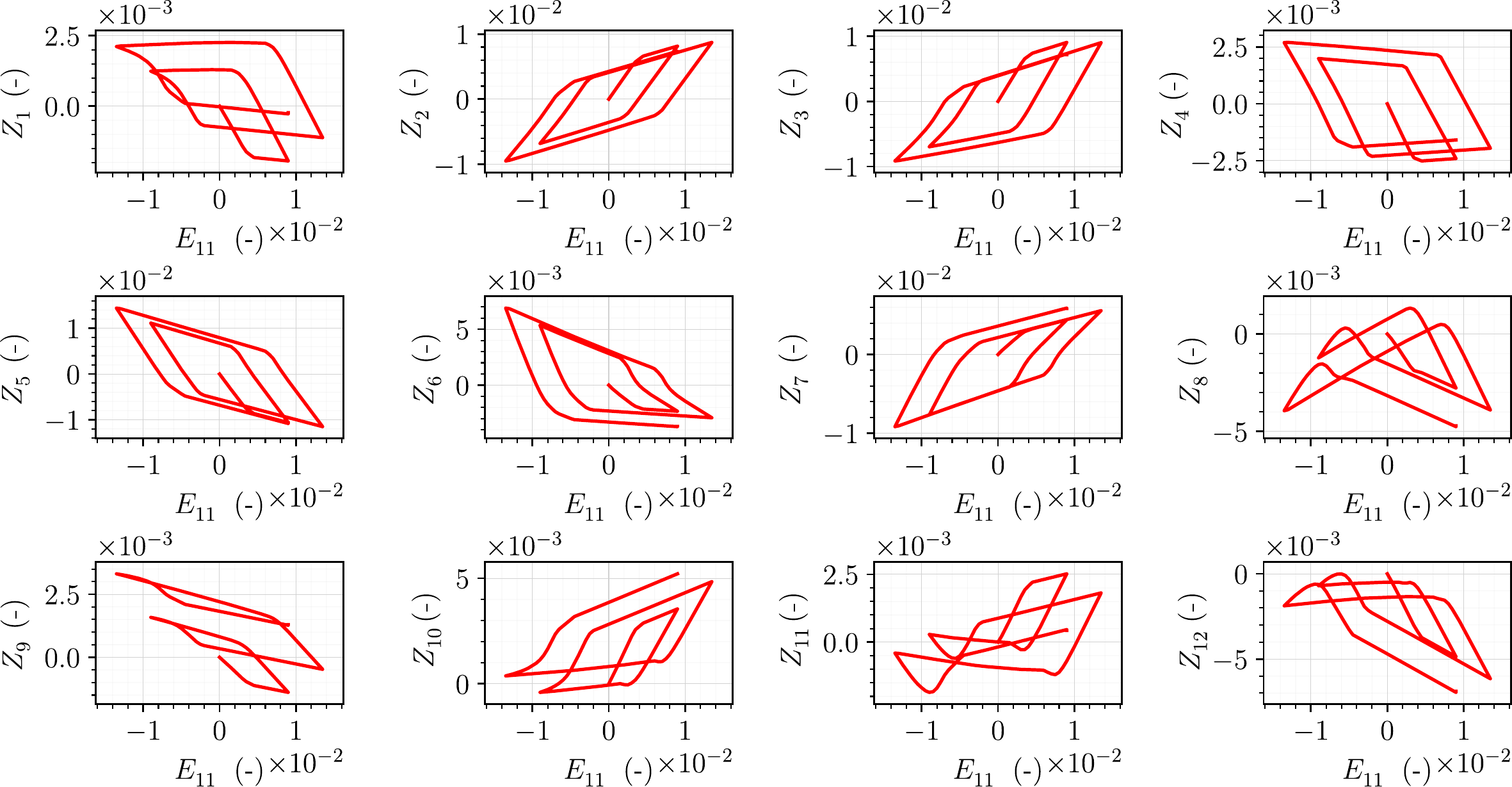}
	\caption{\footnotesize internal state variables in function of strain}
	\label{fig:L6x6_2c}
\end{subfigure}\\ \vspace{0.3cm}
\begin{subfigure}[t]{0.45\textwidth}
  \centering
  \includegraphics[width=\linewidth]{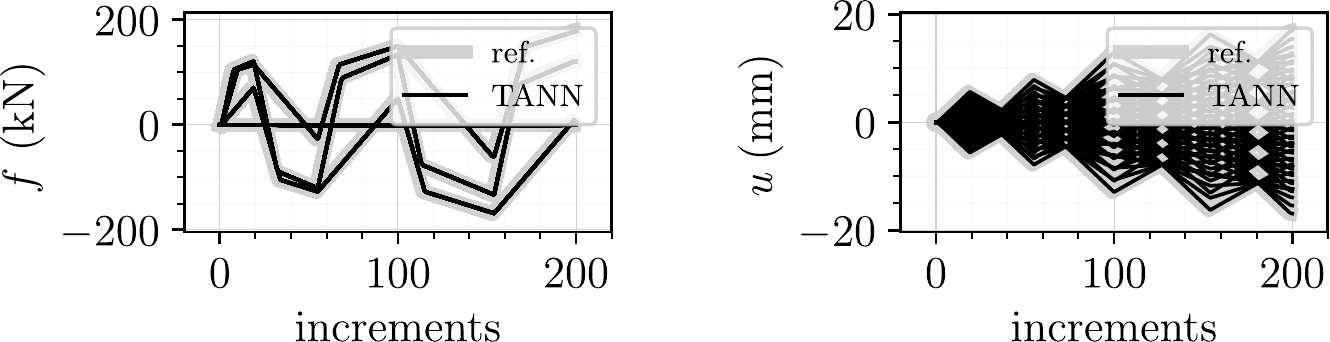}
	\caption{\footnotesize internal coordinates: internal forces $f$ (left) and microscopic displacements $u$ (right)}
	\label{fig:L6x6_2d}
\end{subfigure}
\caption{Response of the $6\times 6$ lattice in terms of stresses (b), internal state variables (c), and internal coordinates (d), for a uni-axial cyclic loading path, with increasing amplitude, and monotonous shearing (a).}
\label{fig:L6x6_2}
\end{figure*}

For the sake of conciseness, we omit to present the predictions for the $4\times 4$ and we show directly the results for the $6\times 6$ lattice. Figure \ref{fig:L6x6_2} displays the material predictions under combined cyclic uni-axial strain path and monotonous shearing. TANN successfully predict the stress-strain behavior, as well as the evolution of the internal coordinates. As far as it concerns the discovered internal state variables, we observe that they differ from those computed for the $2\times 2$ lattice. This is due to the non-uniqueness of thermodynamically admissible internal state variables \citep[cf.][]{maugin1994thermodynamicsA}.

\subsection{2D non-regular lattices}
\begin{figure}[h]
	\centering
	\includegraphics[width=0.5\textwidth]{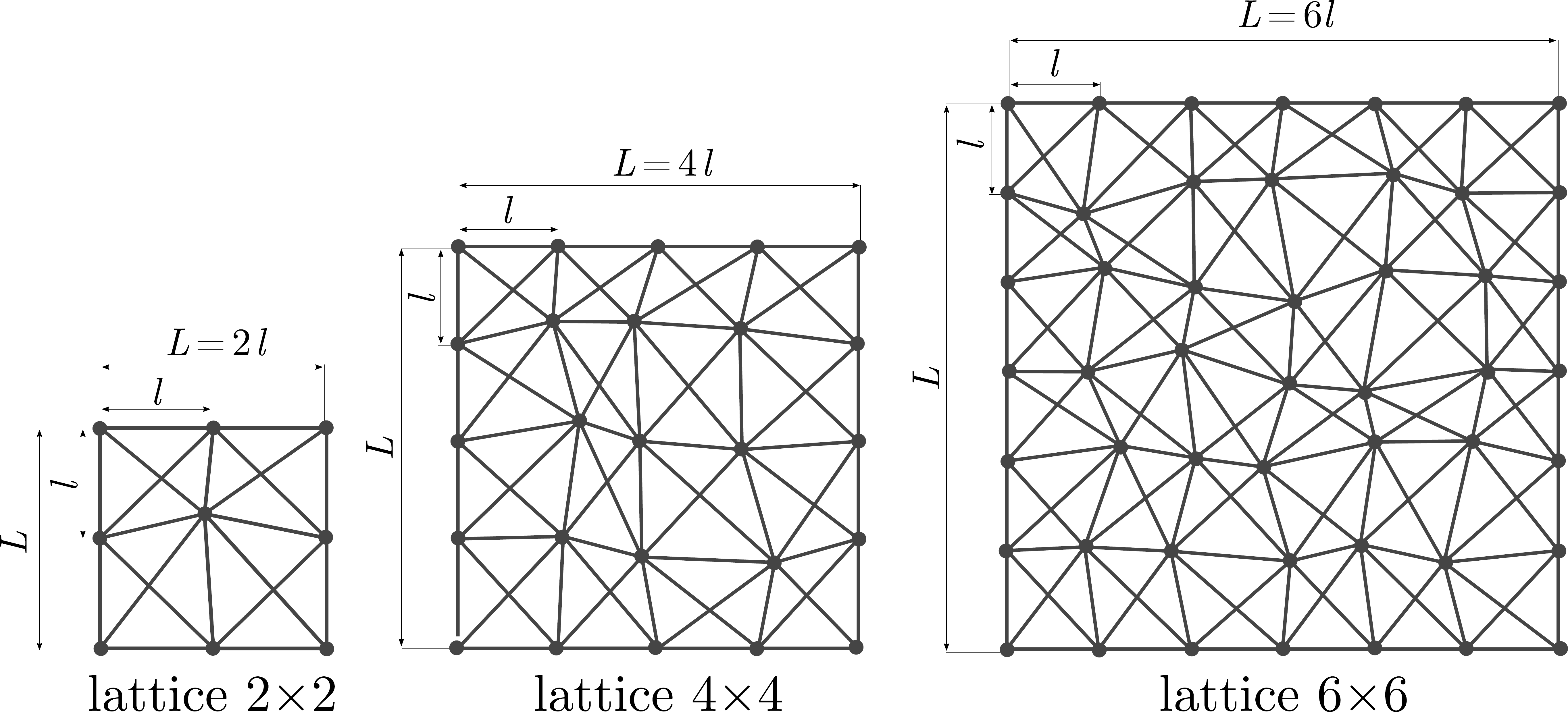}
	\caption{Investigated 2D non-regular lattices.}
	\label{fig:lattice2Dpert}
\end{figure}

\noindent We focus here on non-regular 2D lattices, as shown in Figure \ref{fig:lattice2Dpert}. The coordinates of the lattice structures are perturbed from the regular configuration shown in Figure \ref{fig:lattice2Dreg}, using a normal distribution with zero mean and a standard deviation equal to 0.2. The architecture of the TANN is kept constant in all cases (cf. paragraph \ref{subsec:2Dref}).

\begin{figure*}[h]
\centering
\begin{subfigure}[t]{0.2\textwidth}
  \centering
  \includegraphics[width=\linewidth]{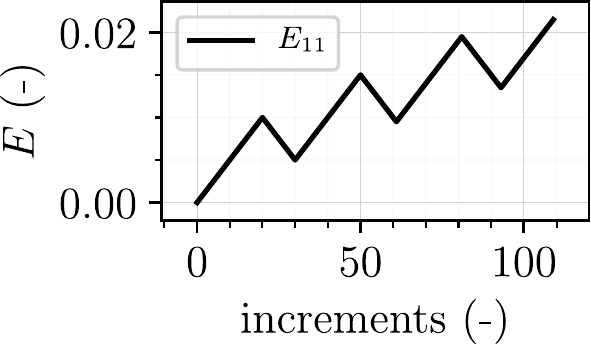}
	\caption{\footnotesize strain path}
	\label{fig:L2x2_PER1a}
\end{subfigure} \hspace{0.3cm}
\begin{subfigure}[t]{0.45\textwidth}
  \centering
  \includegraphics[width=\linewidth]{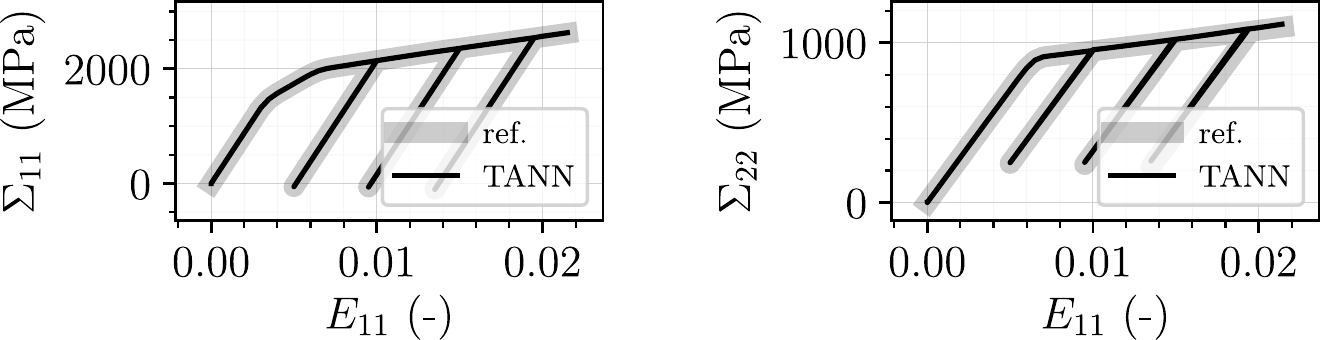}
	\caption{\footnotesize stress-strain behavior}
	\label{fig:L2x2_PER1b}
\end{subfigure}\\ \vspace{0.3cm}
\begin{subfigure}[t]{0.8\textwidth}
  \centering
  \includegraphics[width=\linewidth]{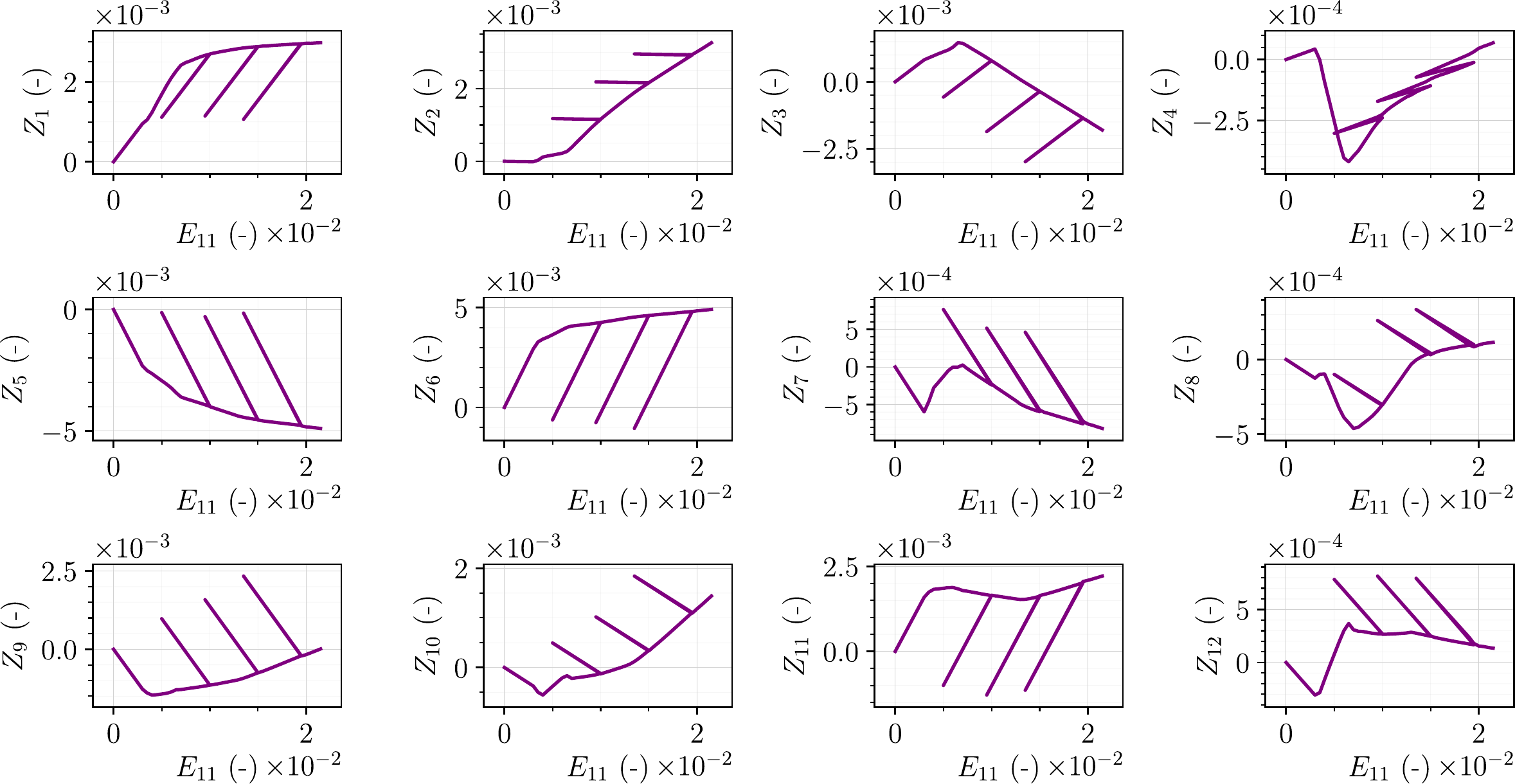}
	\caption{\footnotesize internal state variables in function of strain}
	\label{fig:L2x2_PER1c}
\end{subfigure}\\ \vspace{0.3cm}
\begin{subfigure}[t]{0.45\textwidth}
  \centering
  \includegraphics[width=\linewidth]{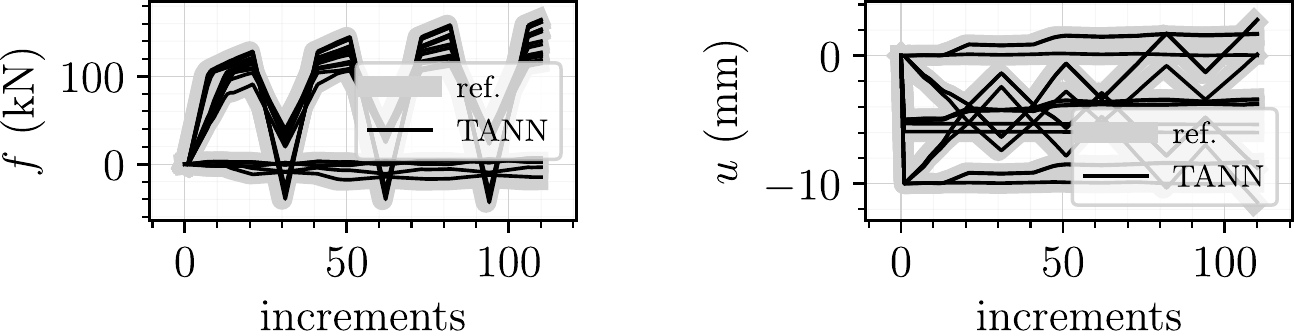}
	\caption{\footnotesize internal coordinates: internal forces $f$ (left) and microscopic displacements $u$ (right)}
	\label{fig:L2x2_2d_PER}
\end{subfigure}
\caption{Response of the $2\times 2$ non-regular lattice in terms of stresses (b), internal state variables (c), and internal coordinates (d), for a uni-axial cyclic loading path (a).}
\label{fig:L2x2_PER}
\end{figure*}

\begin{figure*}[h]
\centering
\begin{subfigure}[t]{0.2\textwidth}
  \centering
  \includegraphics[width=\linewidth]{Figure11_1.pdf}
	\caption{\footnotesize strain path}
	\label{fig:L6x6_PER1a}
\end{subfigure} \hspace{0.3cm}
\begin{subfigure}[t]{0.45\textwidth}
  \centering
  \includegraphics[width=\linewidth]{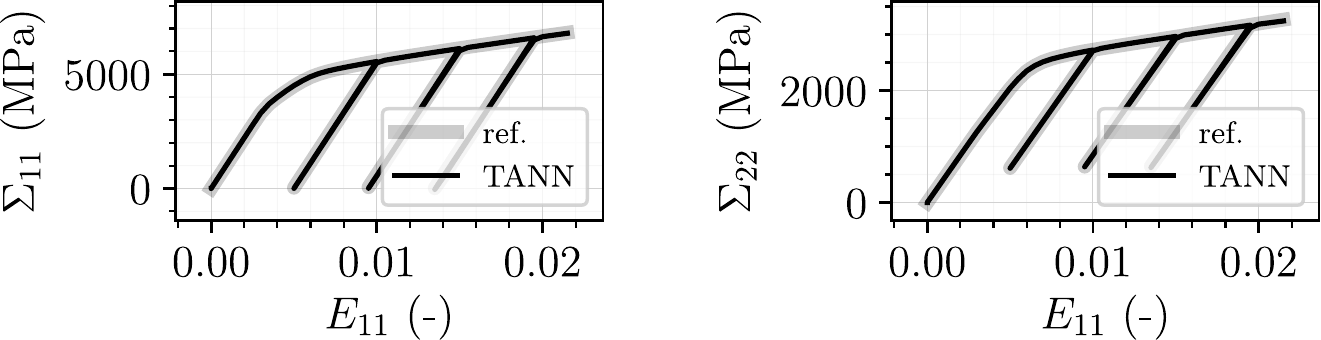}
	\caption{\footnotesize stress-strain behavior}
	\label{fig:L6x6_PER1b}
\end{subfigure}\\ \vspace{0.3cm}
\begin{subfigure}[t]{0.8\textwidth}
  \centering
  \includegraphics[width=\linewidth]{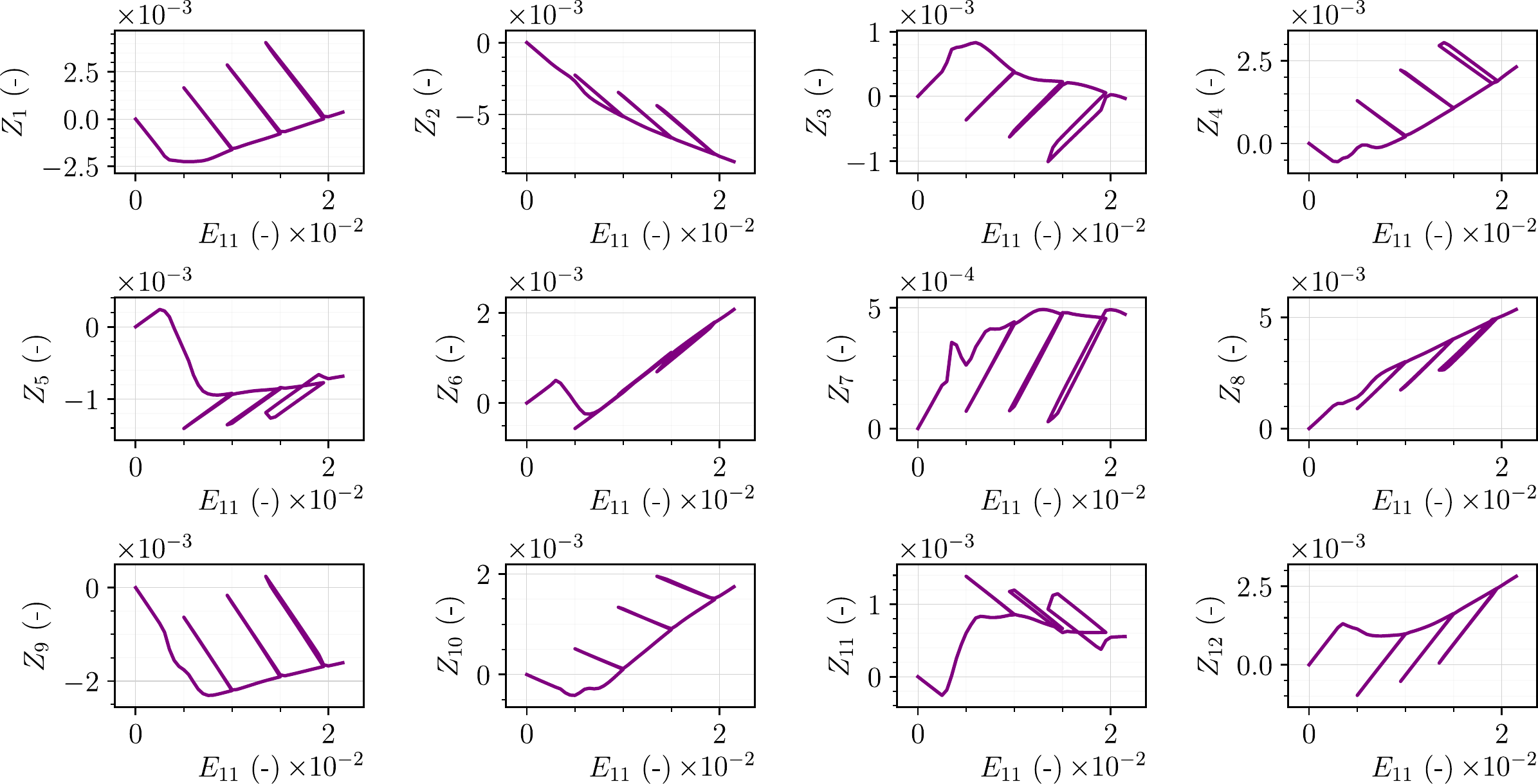}
	\caption{\footnotesize internal state variables in function of strain}
	\label{fig:L6x6_PER1c}
\end{subfigure}\\ \vspace{0.3cm}
\begin{subfigure}[t]{0.45\textwidth}
  \centering
  \includegraphics[width=\linewidth]{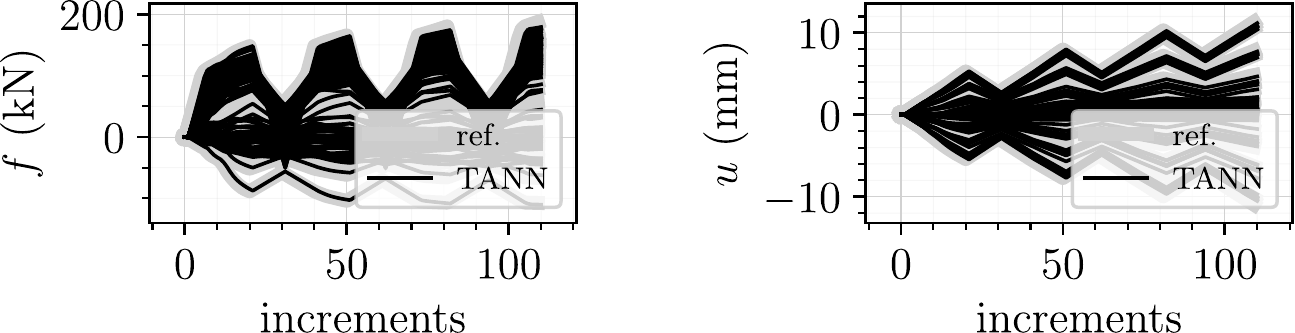}
	\caption{\footnotesize internal coordinates: internal forces $f$ (left) and microscopic displacements $u$ (right)}
	\label{fig:L6x6_2d_PER}
\end{subfigure}
\caption{Response of the $6\times 6$ non-regular lattice in terms of stresses (b), internal state variables (c), and internal coordinates (d), for a uni-axial cyclic loading path (a).}
\label{fig:L6x6_PER}
\end{figure*}

For the sake of conciseness, we only present the results for the $2\times 2$ and $6\times 6$ lattice structures. In both cases, we consider the cyclic uni-axial loading path depicted in Figure \ref{fig:L2x2_PER1a}. Figures \ref{fig:L2x2_PER} and \ref{fig:L6x6_PER} show the comparison between the predictions of TANN and the micromechanical simulations. Despite the non-regularity of the lattice structure (i.e., lack of geometric periodicity), which translates into a higher number of internal state variables, TANN excellently predict the material behavior and the evolution of the internal coordinates.\\
Furthermore, it is worth noticing the dissipative nature of the (discovered) internal state variables. After each unloading (cf. Fig. \ref{fig:L2x2_PER1a}), the internal state variables are always different from zero, due to presence of permanent inelastic deformations, which are perfectly captured by the proposed approach.

\clearpage
\section{Application to double-scale homogenization: the FEM$\times$TANN approach}
\label{sec:app2}
\begin{figure}[h]
	\centering
	\includegraphics[width=0.6\textwidth]{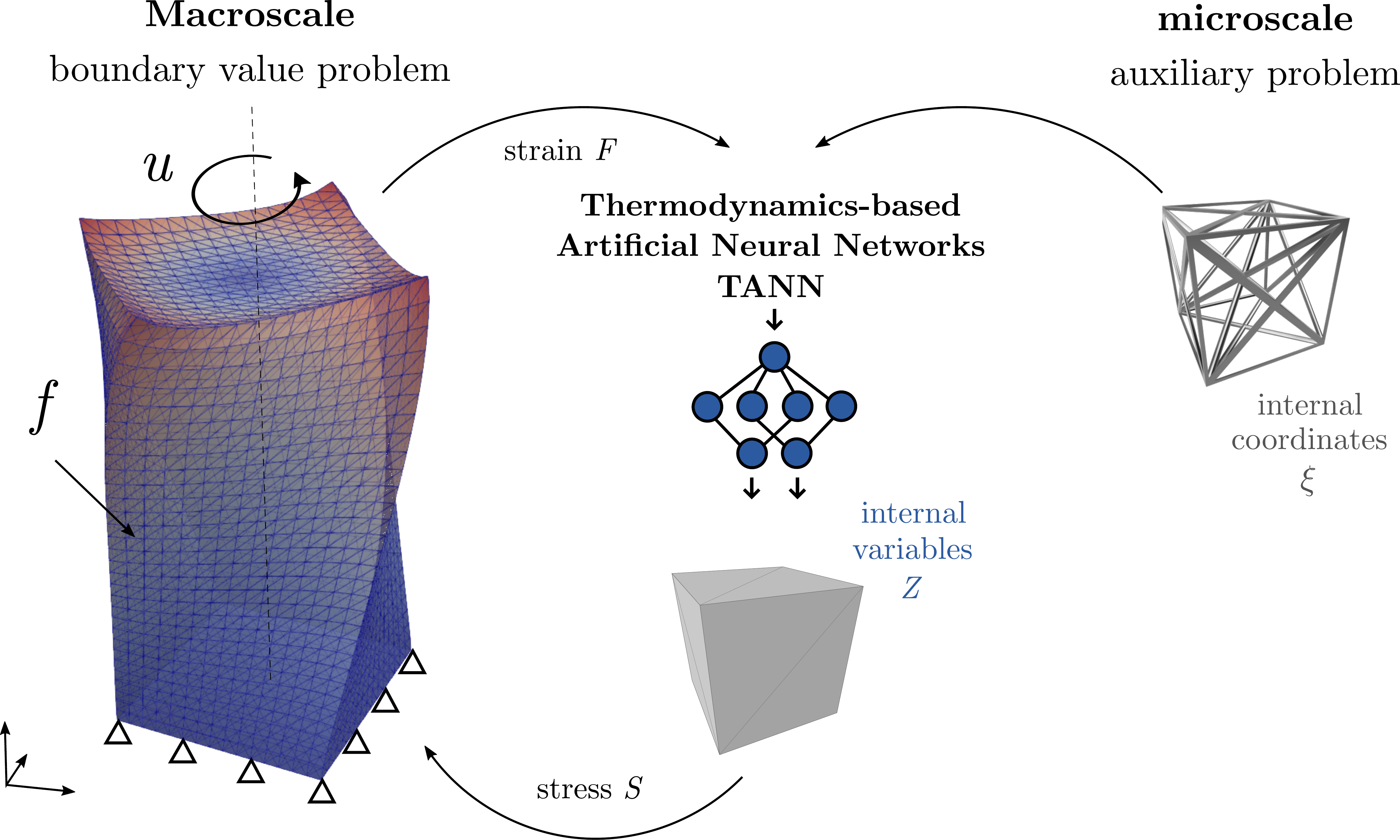}
	\caption{FEM$\times$TANN approach: scheme of the resolution of the large scale boundary value problem with double-scale homogenization and TANN.}
	\label{fig:multiscale}
\end{figure}

\noindent After having demonstrated the capabilities of TANN to predict the behavior of complex lattice materials, we consider large scale, macroscopic lattice structures, whose microstructure is made of spatially (quasi-)periodic distributions of a representative unit-cell $\mathcal{V}$ (cf. the lattice cells presented in Section \ref{sec:app1}). We assume that the scale-separation hypothesis holds, i.e., we assume the existence of two independent scales $x$ and $y=x/\epsilon$, with $\epsilon\ll 1$ being the dimension of the unit-cell. The first scale, $x$, is associated to the macro-scale, while the scale $y$ to the scale of the microstructure (see Fig. \ref{fig:multiscale}). Therefore, asymptotic homogenization can be used \cite{Bakhvalov1989,pinho2009asymptotic}. Notice that the classical linear asymptotic homogenization analysis can be extended to nonlinear problems by means of an incremental formulation \citep[see][]{miehe2002strain}. 

Note that if the assumption of scale separation does not hold, the aforementioned approach cannot be used. This is the case, for instance, of strain localization phenomena taking place at the level of the macroscale. However, a remedy to this issue consists of resorting to higher order continuum theories \citep[see ][, among others]{stefanou2010homogenization,jouan2014using,GODIO2017168,RATTEZ201854,abdallah2020compaction,COLLINSCRAFT2020103975,stathas2022role}. 
Indeed, it is straightforward to broaden the formalism herein adopted by extending the kinematic fields and the dual (in energy) generalized stress using an appropriate upscaling scheme. Accordingly, the solution of the boundary value problem at the macroscale can be separated from that of the auxiliary problem (microscale) resulting in significantly reduced computational costs.

As periodic boundary conditions are used for the generation of the data-sets  for training the extended TANN (see Section \ref{sec:volavg}), our network is trained to predict the solution of the auxiliary problem of homogenization in a wide range of strain increments. Therefore, TANN can be used in recall mode for performing large, multiscale analyses. We refer to this approach as FEM$\times$TANN, to distinguish it from state-of-the-art multiscale approaches (cf. FE\textsuperscript{2}, FEM$\times$DEM).

In the benchmarks presented below the periodic cell used is three-dimensional and is depicted in Figure \ref{fig:multiscale}. Data are generated following the procedure presented in paragraph \ref{subsec:datagen}. TANN architecture is the same as in the 2D cases, but the number of nodes are adjusted to the problem at hand. In particular, the encoder and decoder have 660 nodes, network $NN_Z$ has 88 nodes per layer, while network $NN_{\Psi}$ has 96 nodes. The increase of the nodes is justified by the 3D geometry of the system. The number of the identified thermodynamically admissible ISV is 22 (cf. 12 in 2D). \\

Once TANN are trained, we use them to perform 3D Finite Element analyses (see Fig. \ref{fig:multiscale}) relying on the FEM$\times$TANN approach. The analyses are achieved by a straightforward replacement of classical constitutive models, at the Gauss points, with the trained network. In particular TANN take as inputs the stress and deformation tensors, $\Sigma^t$ and $E^t$, the internal state variables, $Z^t$, at time $t$, and output the updated stress, $\Sigma^{t+\Delta t}$, strain, $E^{t+\Delta t}$, and internal state variables, $Z^{t+\Delta t}$, for an increment of strain, $\Delta E^t$. Energy and dissipation rate are also computed. In implicit analyses, as the ones performed here, the Jacobian $J\triangleq \partial \Delta \Sigma/\partial \Delta E$ is needed as well. This matrix is calculated at each Gauss integration point by virtue of the auto-differentiation of the output $\Delta \Sigma$ with respect to $\Delta E$ \citep{geron2019hands,masiTANNspigl}.\\

For all simulations, the FE models consist of linear tetrahedral elements, with uniform size equal to 0.5 cm, obtained from mesh convergence analyses (see Appendix B). Different loading scenarios are selected: (lcU) uniform compressive deformation; (lcC) uni-axial compressive deformation; (lcT) torsional deformation; and (lcH) history loading path. The results obtained by the FE homogenized model with the TANN-constitutive class are compared with the micromechanical simulations of the entire lattice structure, with different sizes of the unit-cell. In particular, we investigate $\epsilon\in\left[1,1/20\right]$. $\epsilon$ expresses the ratio of the size of the cell over the size of the structure, which is kept constant.

\subsection{Uniform compressive deformation (lcU)}
\noindent First, we consider a cubic lattice structure with height equal to 10 cm. A vertical compressive displacement is applied to the top surface, while the displacements at the bottom boundary are constrained to remain within the horizontal plane. Two cases are considered: (lcU1) vertical displacement equal to 3 cm, elastic case, and (lcU2) equal to 5 cm, inelastic case. These boundary conditions lead to a uniform macroscopic stress field, which is useful for overall testing of the framework and convergence for decreasing $\epsilon$.

\begin{figure*}[h]
\centering
\begin{subfigure}[b]{0.4\textwidth}
  \centering
  \includegraphics[height=0.35\textheight]{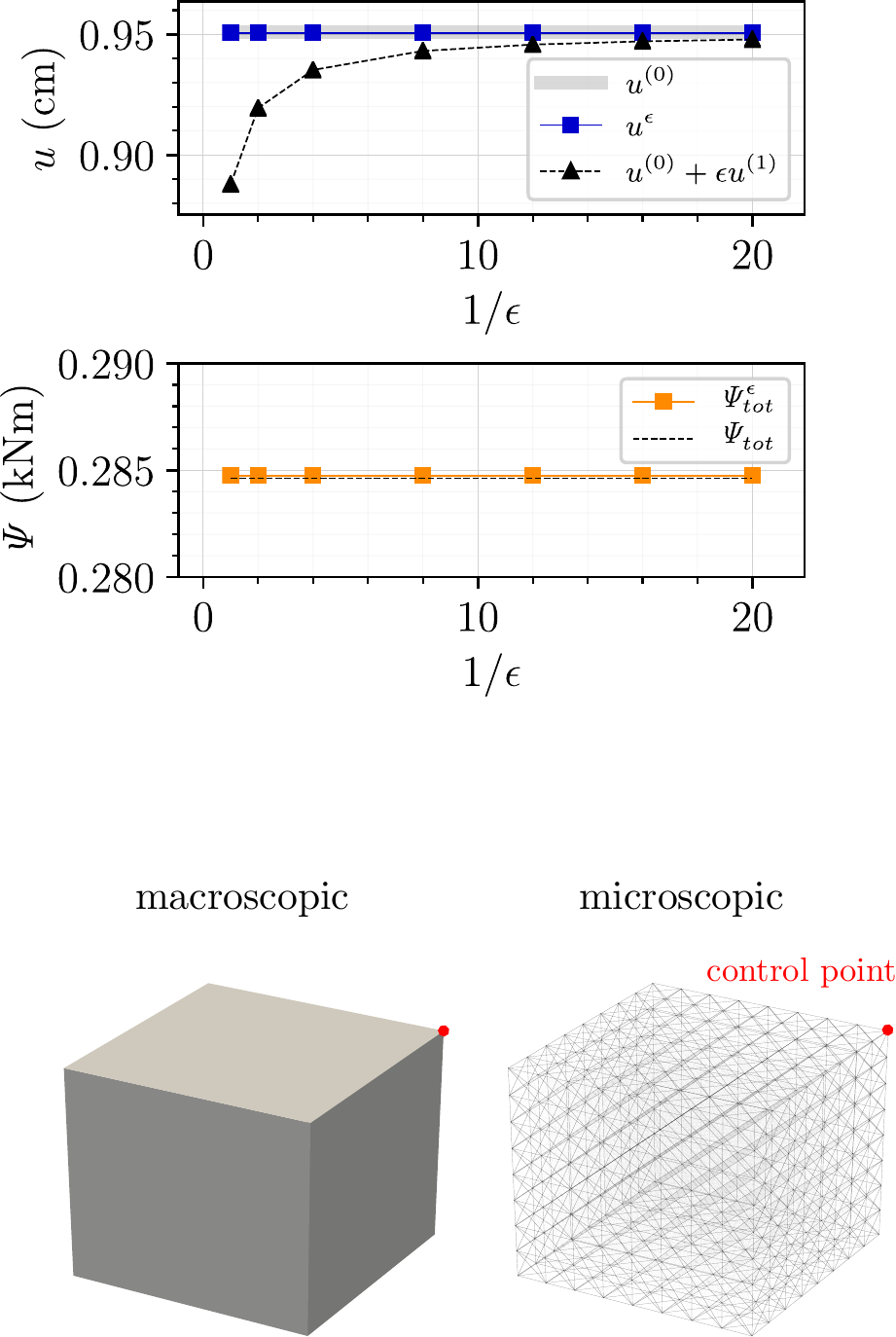}
	\caption{\footnotesize lcU1: vertical displacement equal to 3 cm}
\end{subfigure} \hspace{0.3cm}
\begin{subfigure}[b]{0.4\textwidth}
  \centering
  \includegraphics[height=0.35\textheight]{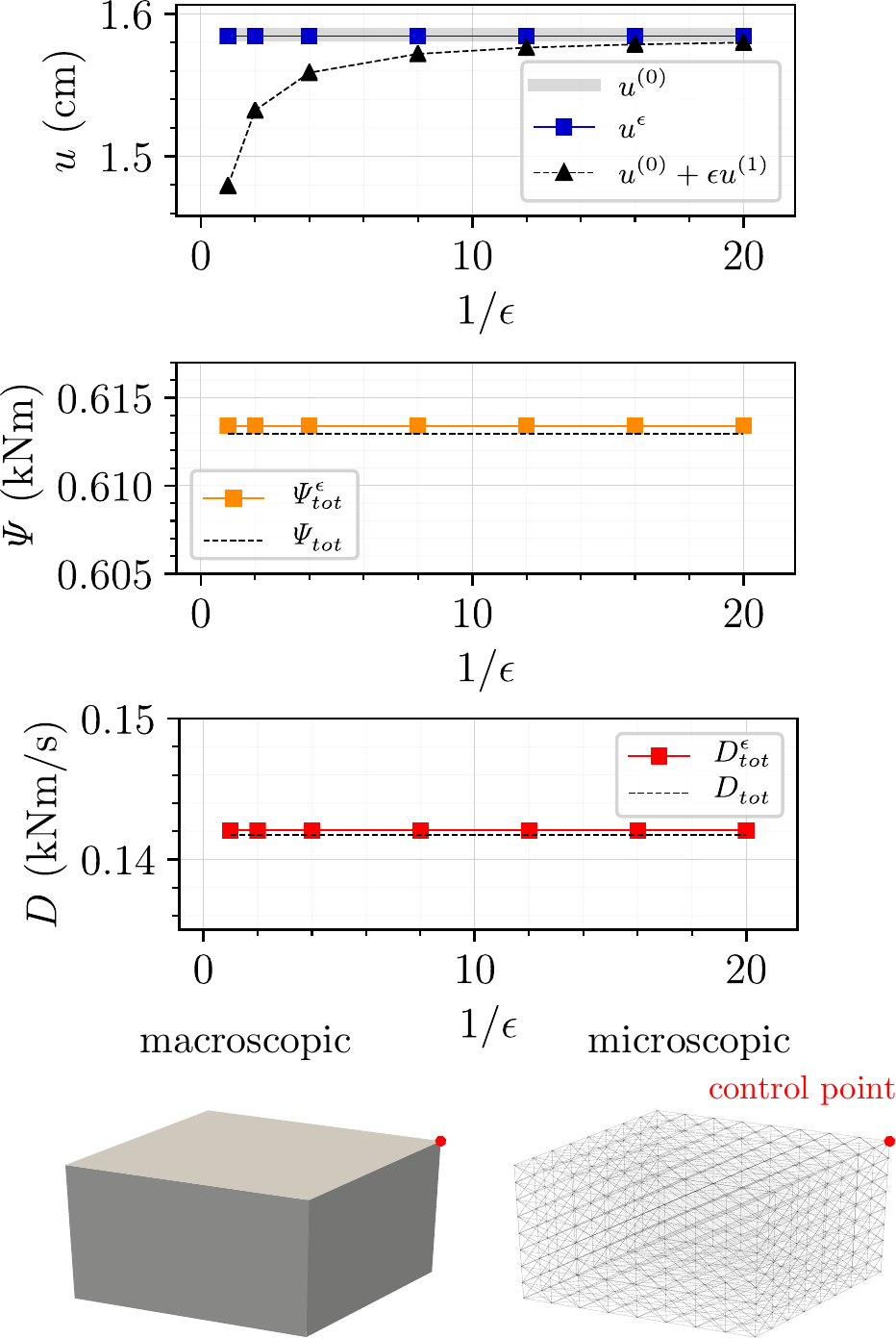}
	\caption{\footnotesize  lcU2: vertical displacement equal to 5 cm}
\end{subfigure}
\caption{Comparison of the displacement at a control point, of the total energy and dissipation between the FE model using TANN and the micromechanical simulations as $\epsilon\rightarrow 0$. On the left, case lcU1: a vertical displacement of 3 cm is applied, leading to uniform elastic deformations ($D=0$). On the right, case lcU2: a vertical displacement of 5 cm is applied, leading to uniform plastic deformations ($D\neq0$).}
\label{fig:hom_uni}
\end{figure*}

The homogenized FE model and the micromechanical simulations are compared in terms of (a) the horizontal displacement at a control point, located at one of the corners of the top boundary (see Fig. \ref{fig:hom_uni}), (b) the total free-energy and (c) dissipation rate. The displacement is computed using the first-order approximation of the microscopic displacement, based on homogenization theory \citep{pinho2009asymptotic}, i.e., $u^{\epsilon} = u^{(0)} +\epsilon u^{(1)}$, with $u^{\epsilon}$ being the displacement obtained from the micromechanical model, $u^{(0)}$ the zeroth-order term$-$that is, the displacement predicted by the FE model, and $u^{(1)}$ the first-order term, which coincides with the displacement internal coordinates, obtained by decoding the internal state variables.

Figure \ref{fig:hom_uni} displays the convergence of the response of the homogenized model based on TANN to the response of the micromechanical simulations, as $\epsilon\rightarrow 0$, both for elasticity and plasticity. For both cases, the error for the displacement predicted by the FE model, at the control point (Figure \ref{fig:hom_uni}), is found to be lower than $0.8\%$, for $\epsilon\leq1/8$, and equal to $0.1\%$, for $\epsilon=1/20$. The total energy and dissipation rate are independent of the unit-cell dimensions, as expected for a uniform deformation loading. In particular, the relative error with respect to the micromechanical simulations for the energy is approximately equal to $0.04\%$ and $0.07\%$ for (lcU1) and (lcU2), respectively. The error in the dissipation rate is $0.12\%$.

By virtue of the decoding of the internal coordinates, the TANN-constitutive class allows to derive not only the first-order approximation of the microscopic displacements, $u^{(1)}$, but additionally the (zeroth-order approximation of the) microscopic stresses, $\sigma^{(1)}$, computed by dividing the internal forces (internal coordinates) with the cross-section of the bars composing the unit-cell. By comparing the microscopic stresses of all the bars connected to the control node of the homogenized model with the micromechanical one, $\sigma^{\epsilon}$, we find a mean error equal to $0.04\%$, for case lcU1, and to $0.02\%$, for case lcU2. The sources of this error can be multiple. It can be attributed to the FE discretization, the tolerances of the Newton-Raphson iterations for solving the nonlinear problem, and the TANN accuracy. In any case, this error is negligible.

As far as it concerns the comparison of the computational cost of the FE simulations using TANN with respect to the micromechanical simulations, we refer to paragraph \ref{subsec:CCR}.

\subsection{Uni-axial compression (lcC)}
\noindent In this scenario, we consider a uni-axial compressive displacement, applied at the top of the structure, and with bottom fixed-end. The same sample of the previous paragraph is used. Two displacements are applied: (lcC1) 2 cm, elastic case, and (lcC2) 5 cm, inelastic case.

\begin{figure*}[h]
\centering
\begin{subfigure}[b]{0.4\textwidth}
  \centering
  \includegraphics[height=0.37\textheight]{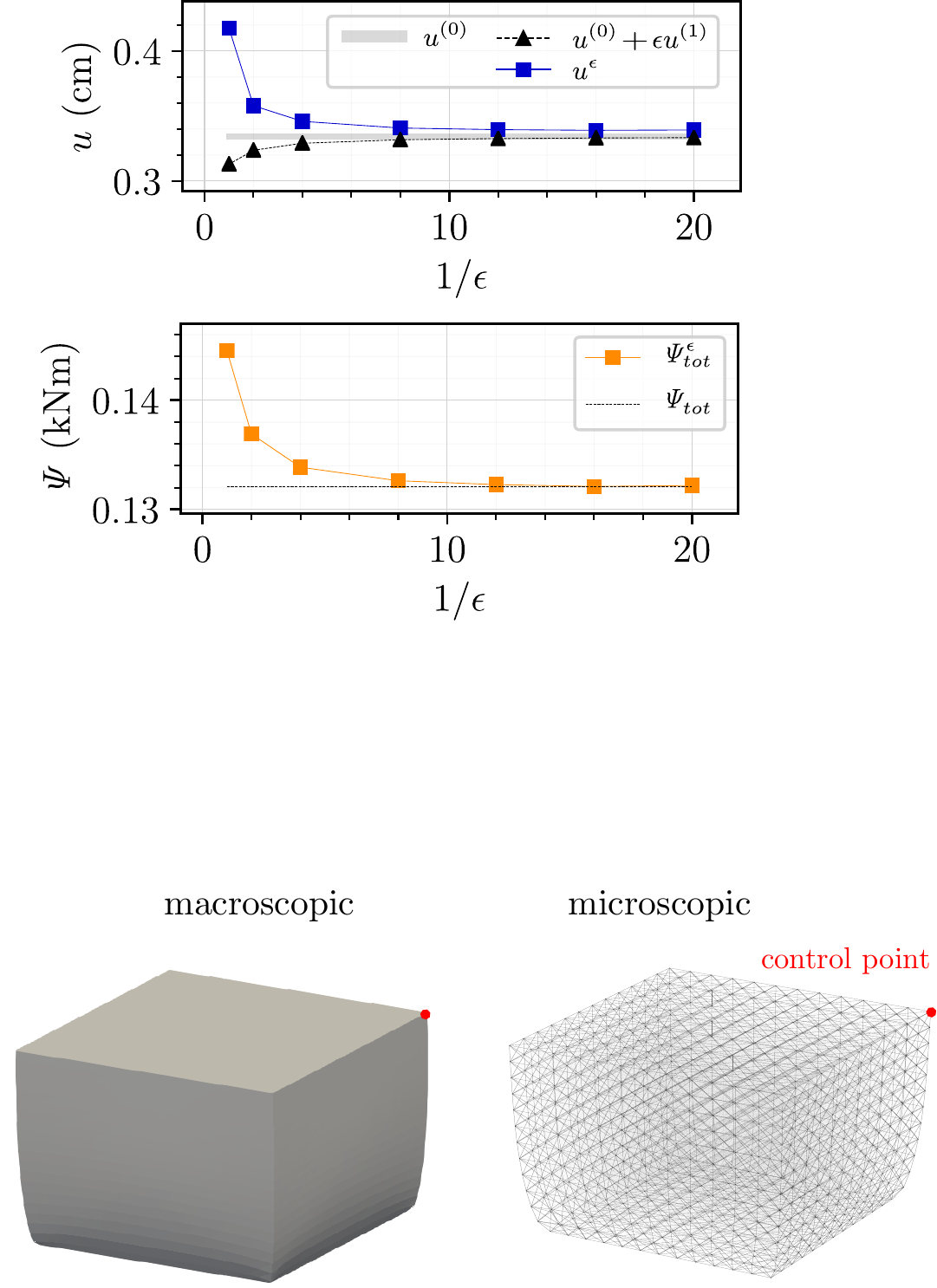}
	\caption{\footnotesize lcC1: vertical displacement equal to 2 cm}
\end{subfigure} \hspace{1cm}
\begin{subfigure}[b]{0.4\textwidth}
  \centering
  \includegraphics[height=0.37\textheight]{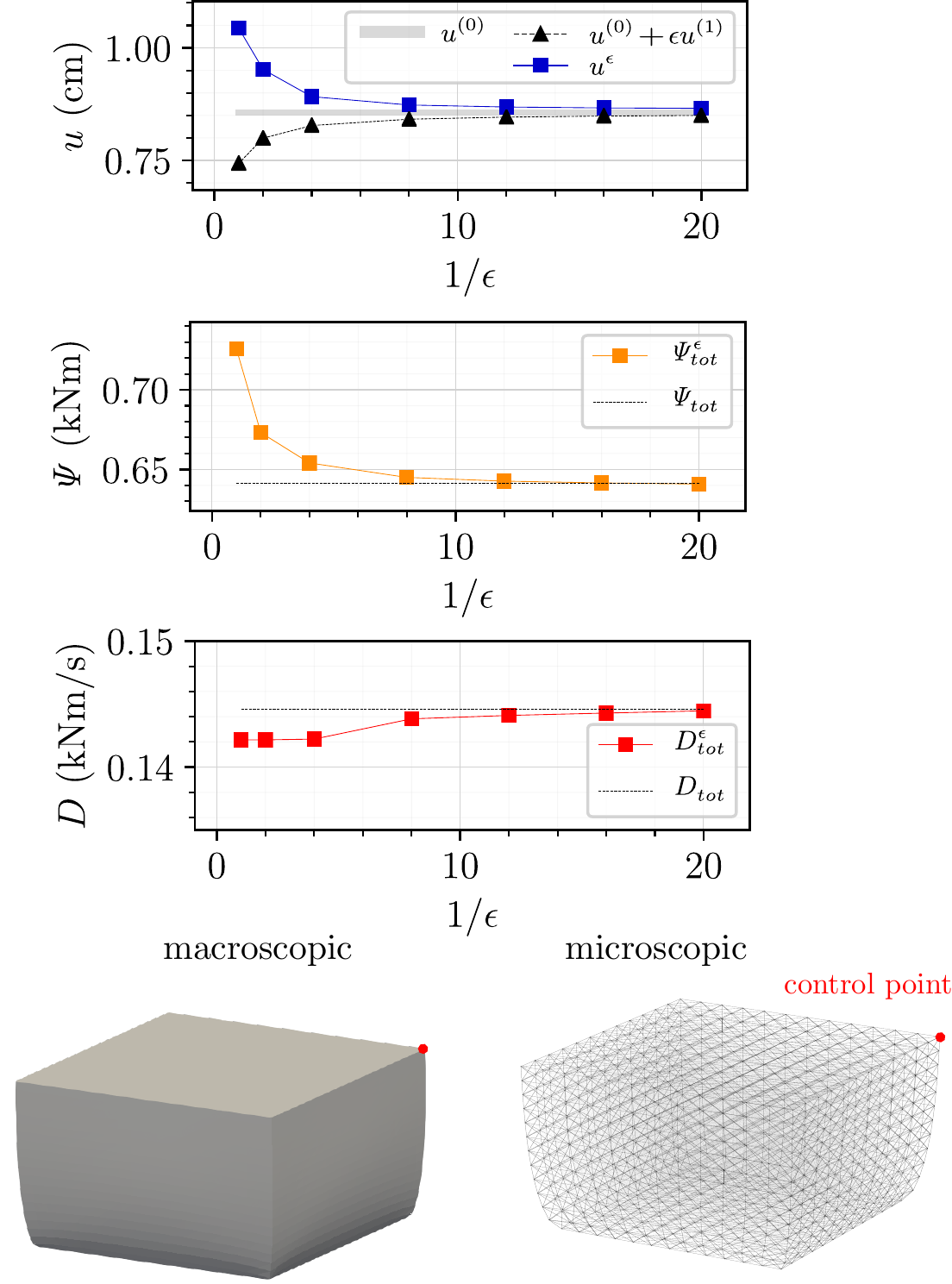}
	\caption{\footnotesize lcC2: vertical displacement equal to 5 cm}
\end{subfigure}
\caption{Comparison of the displacement at a control point, of the total energy and dissipation between the FE model using TANN and the micromechanical simulations as $\epsilon\rightarrow 0$. On the left, case lcC1: uni-axial compression due to a vertical displacement equal to 2 cm, leading to uniform elastic deformations ($D=0$). On the right, case lcC2: uni-axial compression due to a vertical displacement equal to 5 cm, leading to uniform plastic deformations ($D\neq0$).}
\label{fig:hom_comp}
\end{figure*}

\begin{figure*}[h]
\centering
\begin{subfigure}[b]{0.4\textwidth}
  \centering
  \includegraphics[height=0.35\textheight]{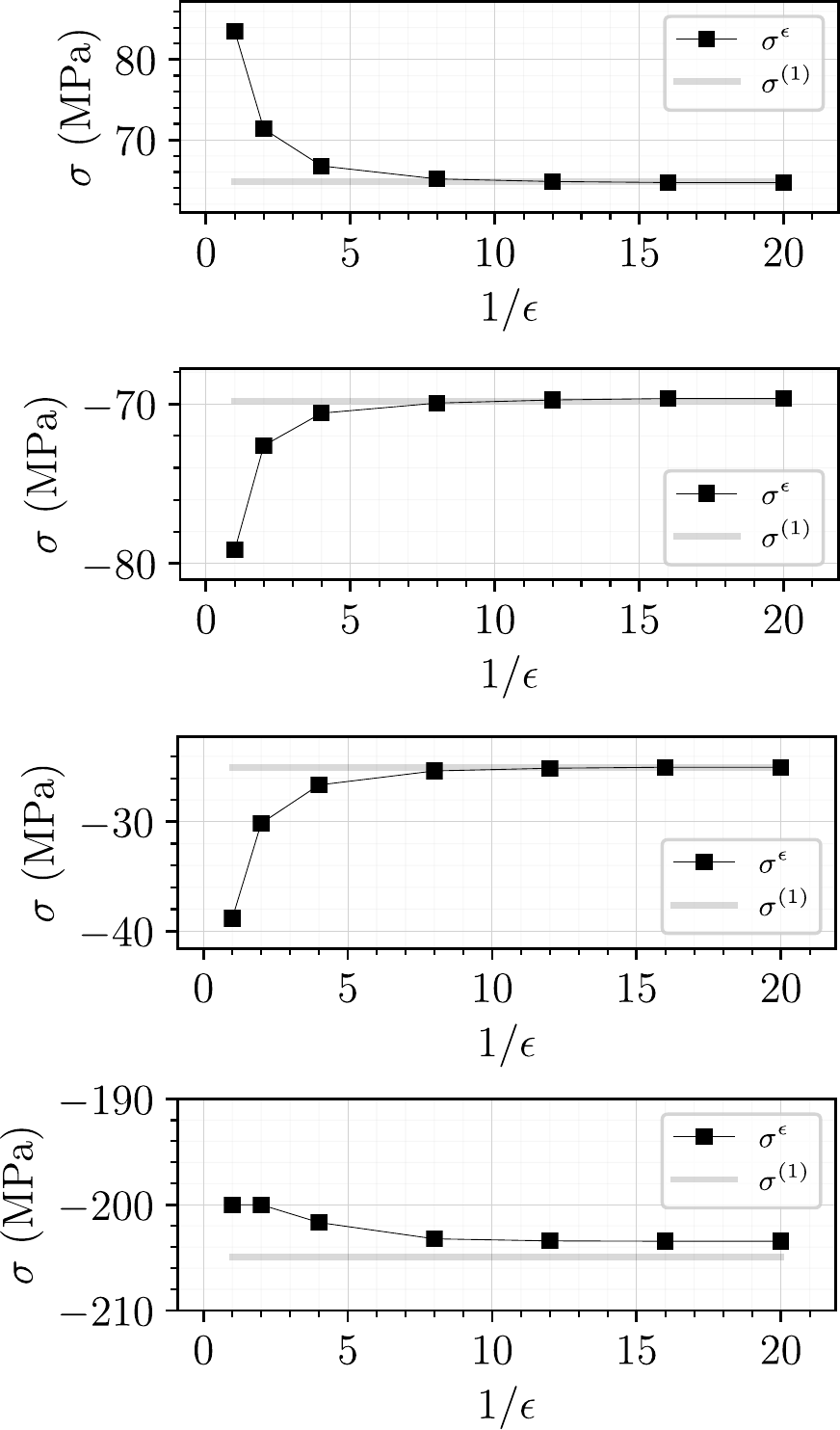}
	\caption{\footnotesize lcC1: vertical displacement equal to 2 cm}
\end{subfigure} \hspace{0.3cm}
\begin{subfigure}[b]{0.4\textwidth}
  \centering
  \includegraphics[height=0.35\textheight]{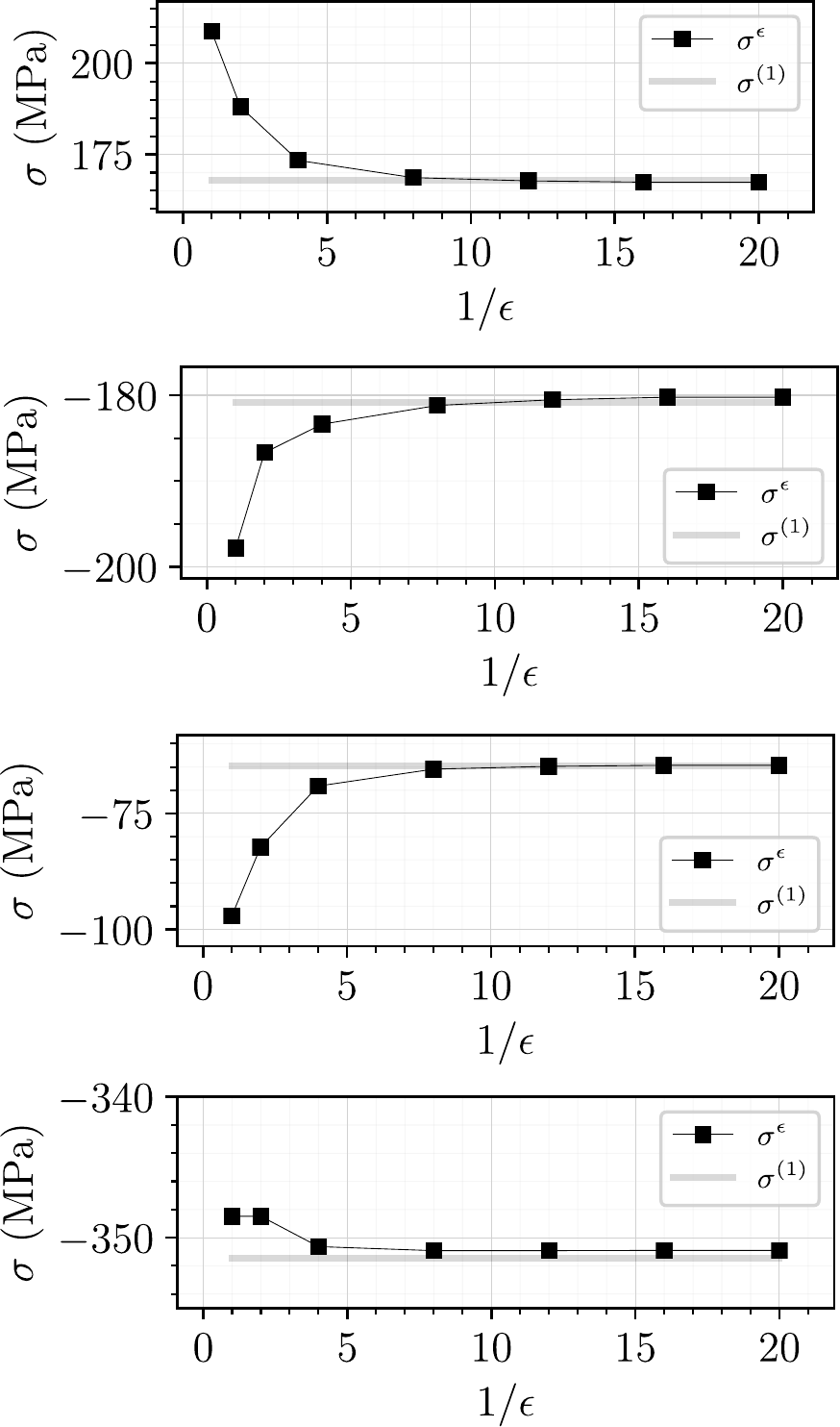}
	\caption{\footnotesize lcC2: vertical displacement equal to 5 cm}
\end{subfigure}
\caption{Comparison of the stresses of all bars connected to the control point (Figure \ref{fig:hom_comp} between the FE model using TANN and micromechanical simulations for case lcC, that is uni-axial compression due to a vertical displacement equal to 2 cm (lcC1) and 5 cm (lcC2).}
\label{fig:hom_comp_stress}
\end{figure*}

Figures \ref{fig:hom_comp} and \ref{fig:hom_comp_stress} show the comparison between the homogenized model and the micromechanical one. TANN allow to accurately predict the micromechanical response.
In particular, we find the following relative errors for $\epsilon=1/20$: $1.6\%$ for the horizontal displacement at the control point, $u^{\epsilon}$ and $u^{(0)}+\epsilon u^{(1)}$, $0.05\%$ for the total energy, $\Psi_{tot}^{\epsilon}$ and $\Psi_{tot}$, $0.1\%$ for the total dissipation, $D_{tot}^{\epsilon}$ and $D_{tot}$, and $0.35\%$ for the microstresses, $\sigma^{\epsilon}$ and $\sigma^{(1)}$.

Contrary to the previous case (lcU), we report displacements in a region of high strain and stress gradients (inside the boundary layer). It is well known that asymptotic homogenization at the first-order cannot capture in detail such effects. However, the errors are very limited for the application at hand, showing the efficiency of TANN even in inelasticity. 

\subsection{Torsional deformation (lcT)}
\noindent We consider a lattice structure with width and depth equal to 10 cm and height equal to 20 cm. A torsional displacement is applied at the top boundary, leaving free all the others degrees of freedom in order to allow warping. All degrees of freedom at the bottom are blocked. Two cases are considered: torsion (lcT1) with an angle $\omega=40^{\circ}$, and (lcT2)  with $\omega=60^{\circ}$. 

\begin{figure*}[h]
\centering
\begin{subfigure}[b]{0.4\textwidth}
  \centering
  \includegraphics[height=0.5\textheight]{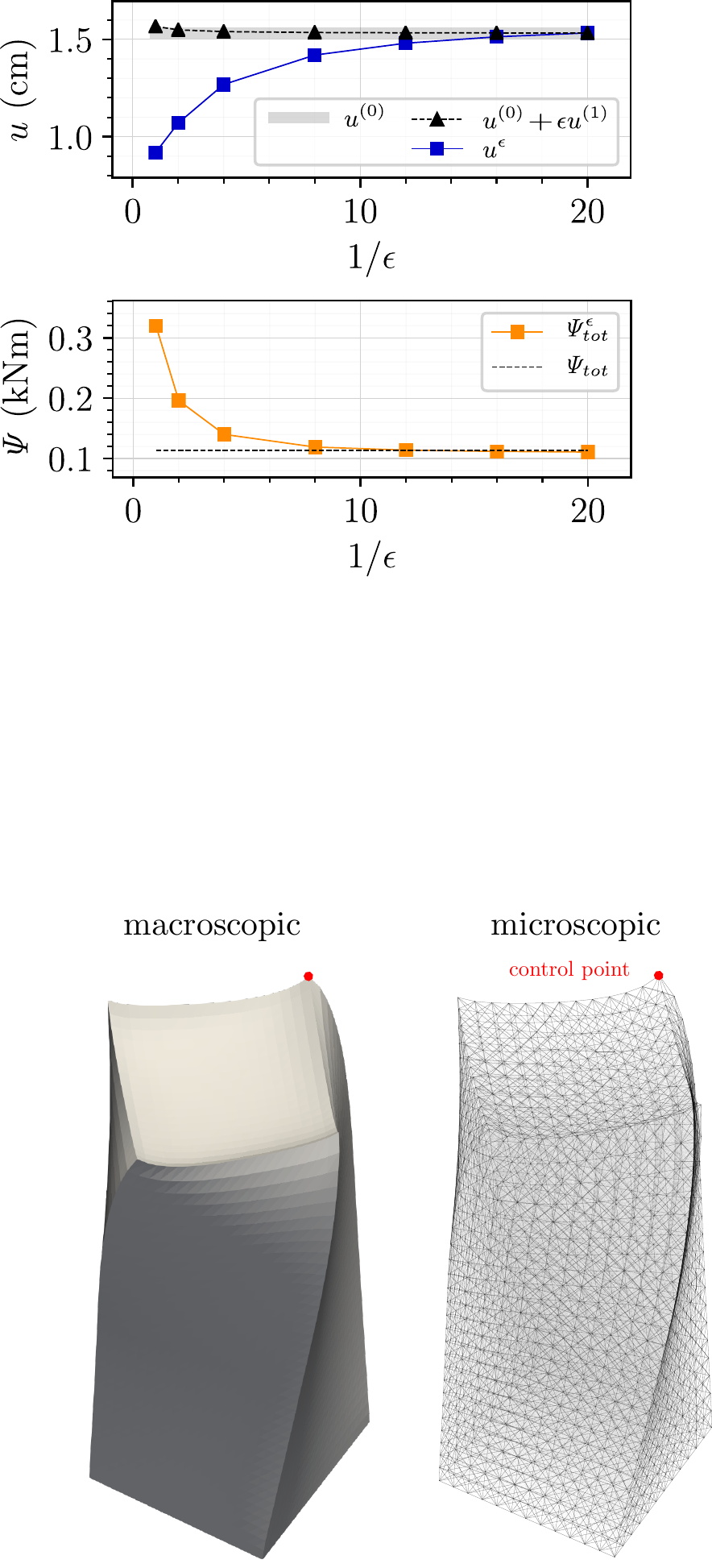}
	\caption{\footnotesize lcT1: $\omega=40^{\circ}$}
\end{subfigure} \hspace{0.3cm}
\begin{subfigure}[b]{0.4\textwidth}
  \centering
  \includegraphics[height=0.5\textheight]{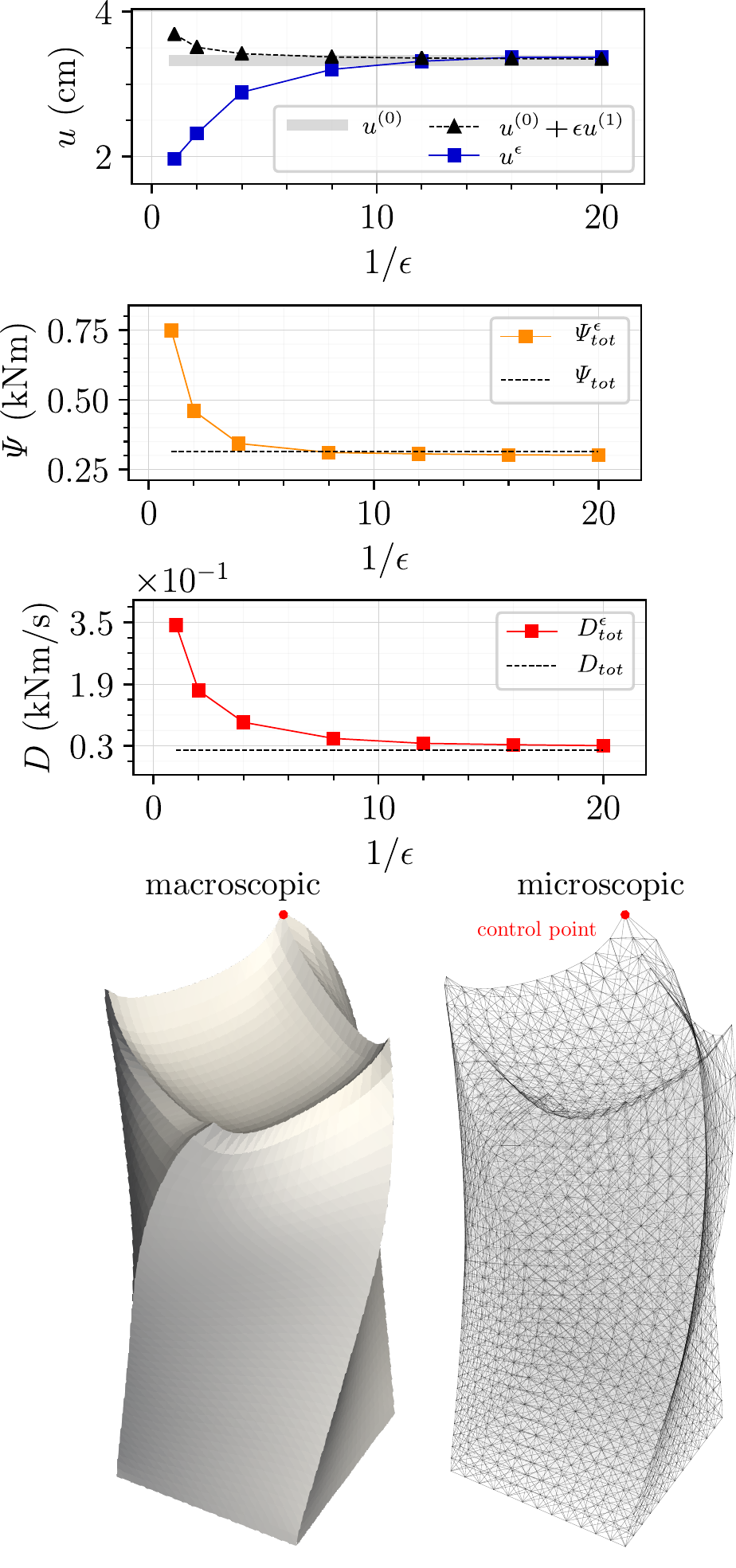}
	\caption{\footnotesize lcT2: $\omega=60^{\circ}$}
\end{subfigure}
\caption{Comparison of the displacement at a control point, of the total energy and dissipation between the FE model using TANN and the micromechanical simulations as $\epsilon\rightarrow 0$. On the left, case lcT1: torsional deformation due to an angle $\omega=40^{\circ}$, leading to uniform elastic deformations ($D=0$). On the right, case lcT2: torsional deformation due to an angle $\omega=60^{\circ}$, leading to uniform plastic deformations ($D\neq0$).}
\label{fig:hom_tors}
\end{figure*}

Figure \ref{fig:hom_tors} shows the comparison between the homogenized model and the micromechanical one. An overall good agreement is found. For $\omega=40^{\circ}$ and $\epsilon=1/20$, the error with respect to the micromechanical simulations is $0.03\%$ in the vertical displacement, at the control point and $1.3\%$ in total energy. For $\omega=60^{\circ}$ and $\epsilon=1/20$, the error in the displacements is $0.68\%$, $1.9\%$ in the total energy and $4.2\%$ in the dissipation rate. The relatively high errors in energy and dissipation, with respect to previous loading scenarios (lcU and lcC), are due to the presence of warping and to the fact that boundary layer effects are not negligible. Indeed, at the top boundary, we observe stress gradients whose wavelength has amplitude comparable to the size of the unit-cell \citep{Bakhvalov1989}, see deformed shapes in Figure \ref{fig:hom_tors}.\\
\indent It is worth noticing that, despite the limit of validity of the asymptotic homogenization theory (at first-order), the FE model and the TANN-constitutive class give excellent results with respect to the micromechanical simulations, especially for the displacements, whose approximation, differently from stresses, is of the first-order. Remedies to the boundary layer effects can however be applied to the proposed methodology \citep[see][]{Bakhvalov1989}.

\subsection{Cyclic loading (lcH)}
\label{subsec:lcH}
\begin{figure*}[h]
\centering
  \includegraphics[width=\textwidth]{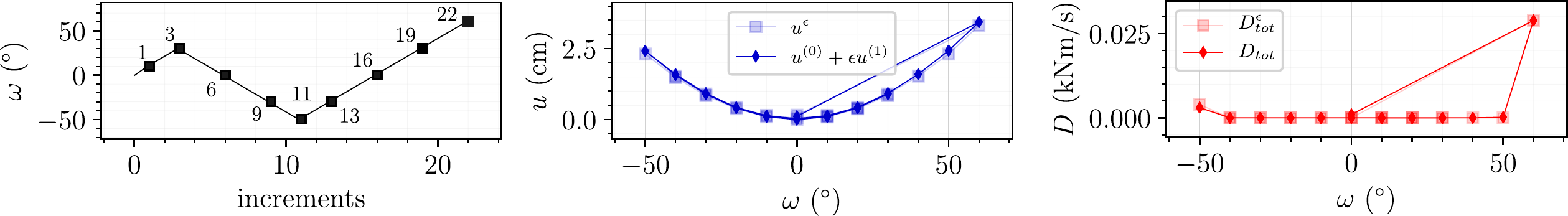} 
\caption{Comparison between the homogenized and micromechanical model, in terms of the first-order approximation of the vertical component of the displacement field at the control point (see Fig. \ref{fig:hom_tors}) and the zeroth-order approximation of the total dissipation rate.}
\label{fig:cyc}
\end{figure*}

\noindent We consider, in this last loading scenario, a cyclic torsional deformation paths of increasing amplitude, see Figure \ref{fig:cyc}. The final, unloading increment consists of releasing the displacements at the top boundary, from $\omega=60^\circ$. This unloading allows to observe permanent displacements due to plasticity. Boundary conditions and geometry are kept the same as in the case lcT. 

Figure \ref{fig:cyc} shows the results of the homogenized and micromechanical models, in terms of the first-order approximation of the vertical displacement at the control point (see Fig. \ref{fig:hom_tors}) and the zeroth-order approximation of the total dissipation rate. It is clear that the homogenized model using TANN agrees very well with the micromechanical solution, despite the demanding loads applied.

Particularly interesting is the comparison at the final increment, where a sudden unloading of the top boundary displacement is performed. The homogenized model accurately predicts the residual, non-zero vertical displacement at the control point and dissipation rate.

\begin{figure*}[h]
\centering
\begin{subfigure}[b]{0.49\textwidth}
  \centering
  \includegraphics[width=\textwidth]{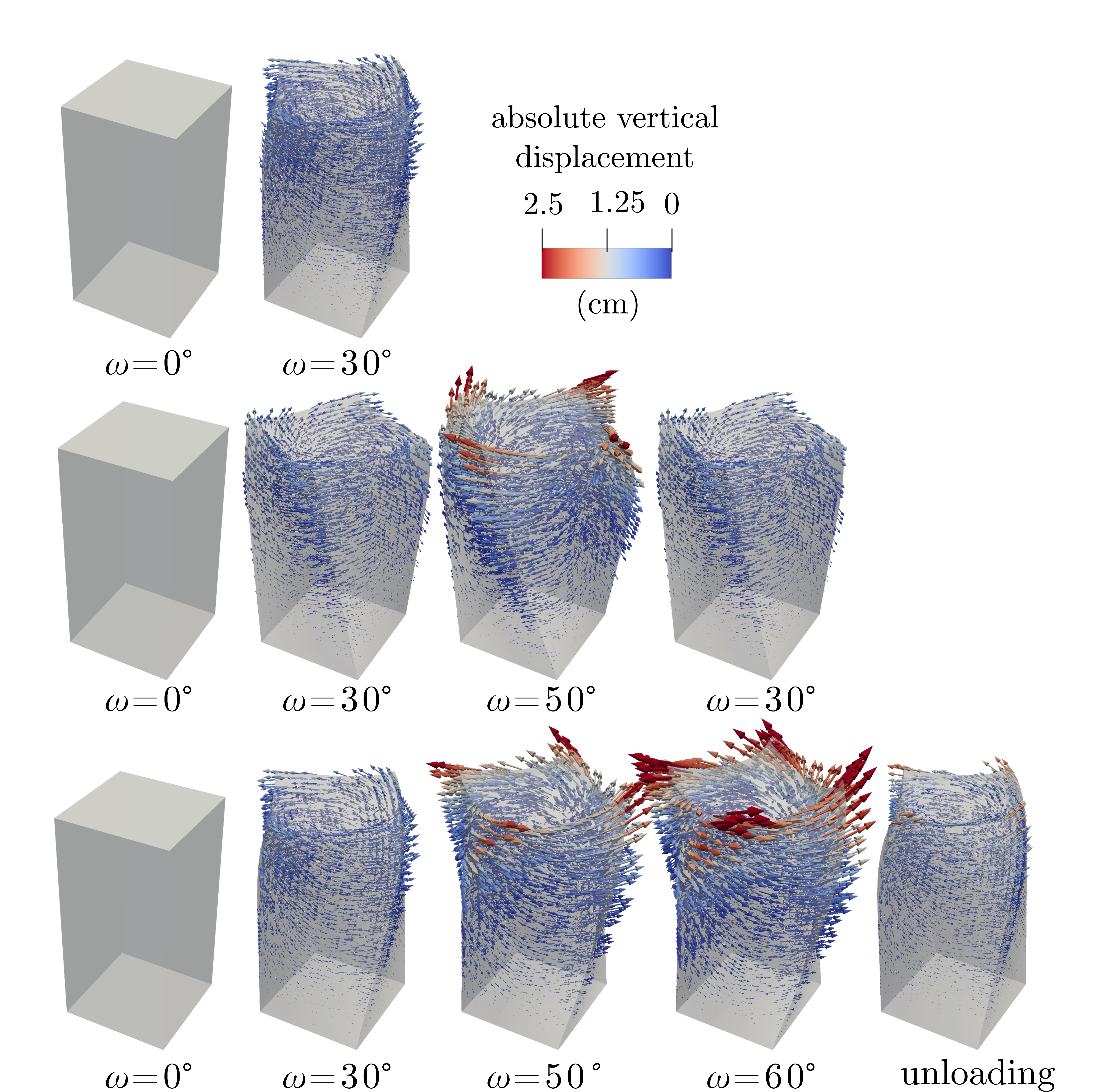}
  \caption{\footnotesize vertical component of the displacement field}
\end{subfigure}
\begin{subfigure}[b]{0.49\textwidth}
  \centering
  \includegraphics[width=\textwidth]{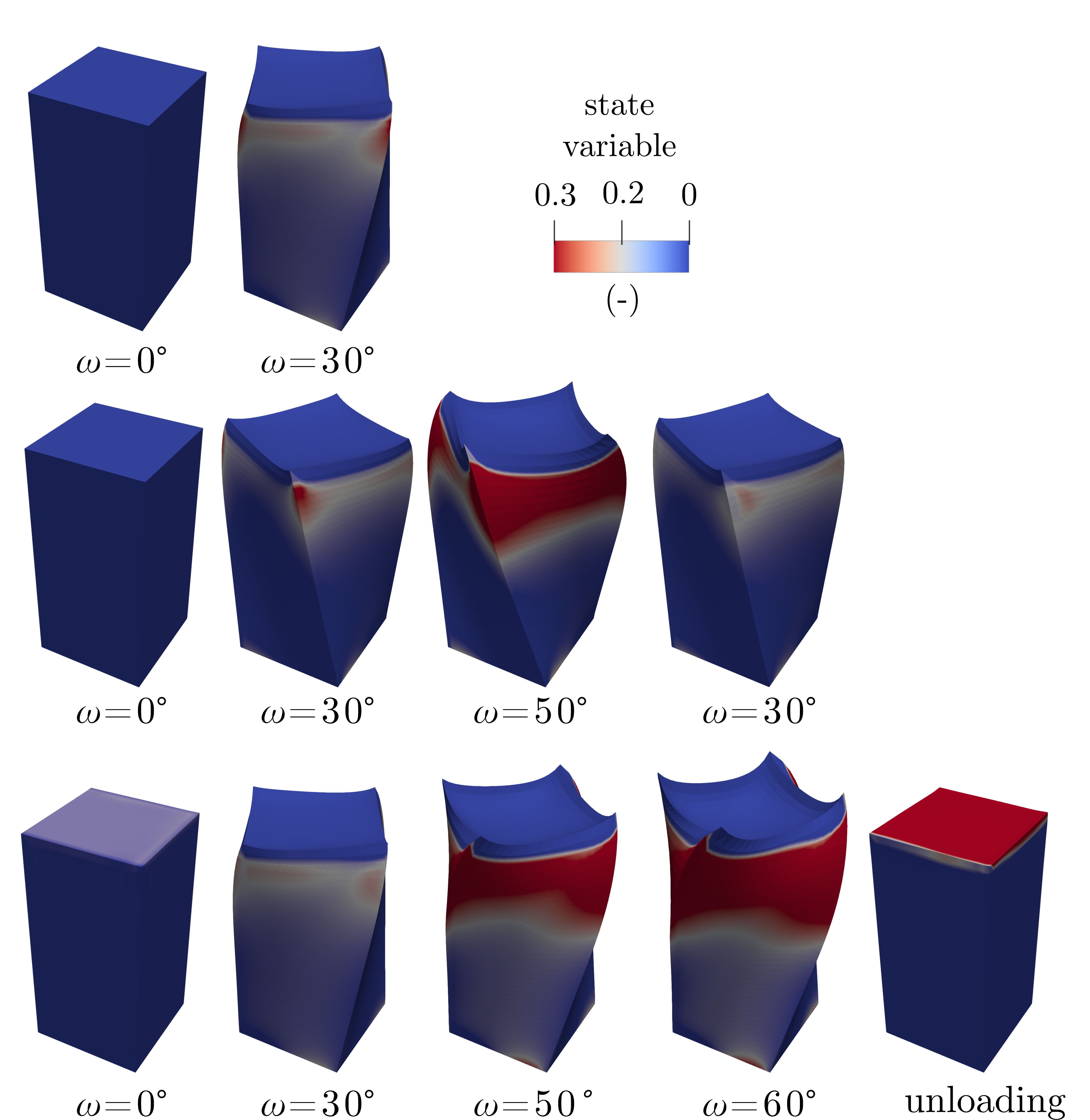}
  \caption{\footnotesize internal state variable}
\end{subfigure}
\caption{Evolution of the vertical component of the displacement field (a) and one of the 22 internal state variables (b) during the FE analyses for the cyclic loading path at different steps.}
\label{fig:cyc_ISV}
\end{figure*}

Figure \ref{fig:cyc_ISV} shows the evolution of the vertical component of the displacement field and that of one of the internal state variables, within the homogenized structure. It is worth noticing that, after the unloading, the ISV are different from zero. This is a direct consequence of the dissipative nature of the internal state variables, identified with the proposed methodology. Furthermore, as discussed in Section \ref{sec:app1}, each internal variable is representative of a particular stress-strain configuration at the microscopic level. Feature extraction methods may further be used to investigate deeper the physical nature of the ISV.

\subsection{Computational cost}
\label{subsec:CCR}
\noindent To conclude the comparison between the FE model based on TANN against the (exact) micromechanical simulations, it is interesting to compare the associated computational cost.
The comparison is performed in terms of calls to the constitutive functions (bars, in the micromechanical model, and Gauss points, in the FE model), during the loading. For making the comparison easier, we define the Computational Cost Ratio (CCR). The choice for this comparison is based on several reasons. First of all, we want a measure independent of the computer system, of the ANN platform used (Tensorflow), and of the optimization of the finite and lattice codes used. Therefore, computational time cannot be used for comparisons.
Second, different constitutive relationships may be used at the microscopic level, which may demand higher or smaller computational time in the lattice code.

\begin{figure*}[h]
\centering
\begin{subfigure}[b]{0.45\textwidth}
  \centering
  \includegraphics[width=0.75\textwidth]{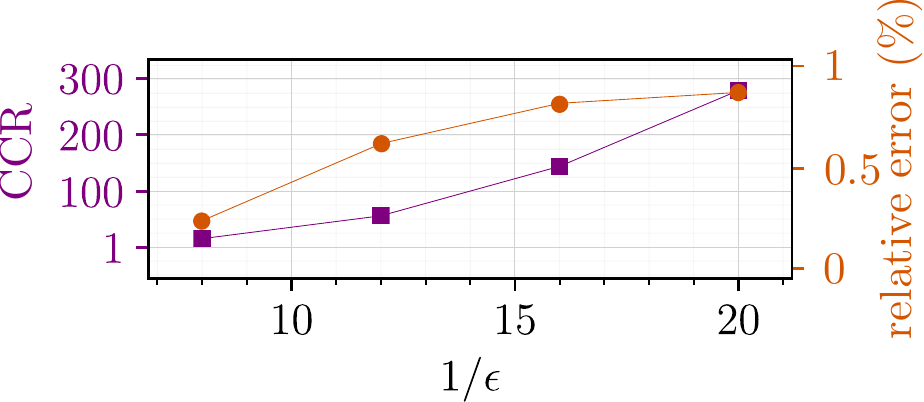}
	\caption{\footnotesize uni-axial compression, coarse FE mesh}
	\label{fig:cost_homa}
\end{subfigure}
\begin{subfigure}[b]{0.45\textwidth}
  \centering
  \includegraphics[width=0.75\textwidth]{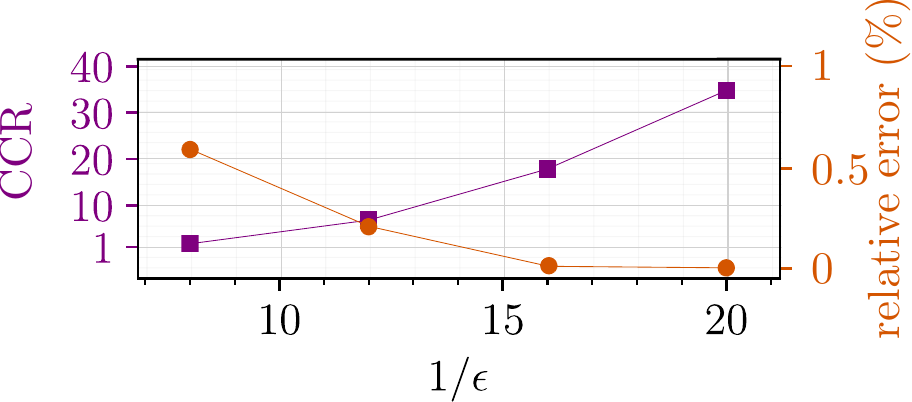}
	\caption{\footnotesize uni-axial compression, fine FE mesh}
	\label{fig:cost_homb}
\end{subfigure}\\ \vspace{0.3cm}
\begin{subfigure}[b]{0.45\textwidth}
  \centering
  \includegraphics[width=0.75\textwidth]{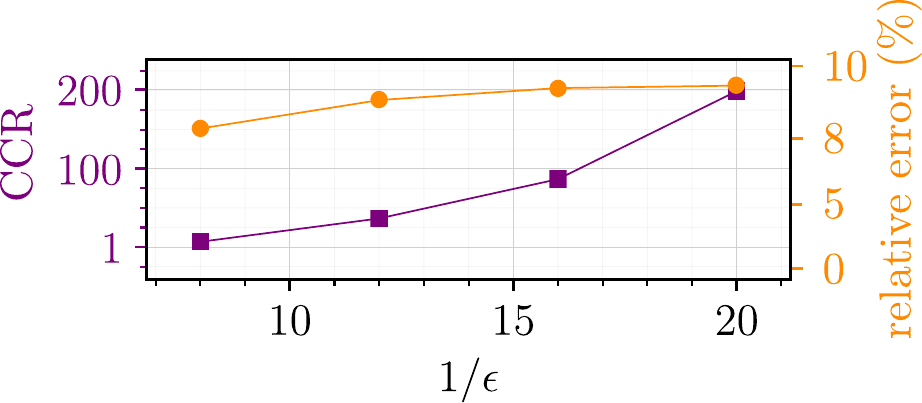}
	\caption{\footnotesize torsional deformation, coarse FE mesh}
	\label{fig:cost_homc}
\end{subfigure} 
\begin{subfigure}[b]{0.45\textwidth}
  \centering
  \includegraphics[width=0.75\textwidth]{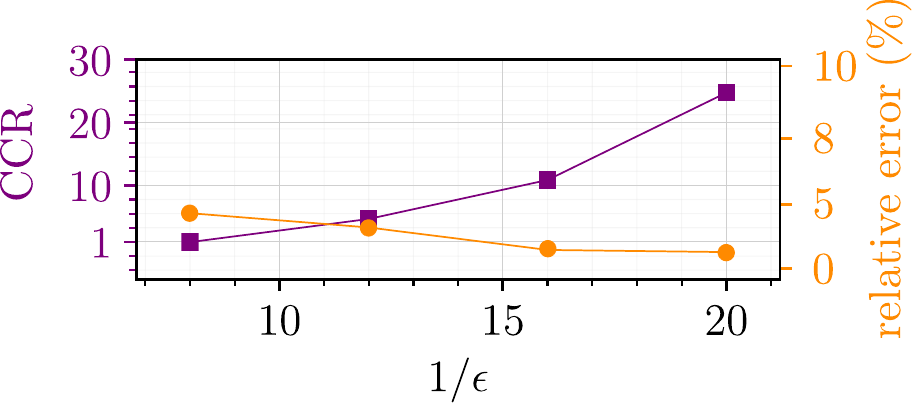}
	\caption{\footnotesize torsional deformation, fine FE mesh}
	\label{fig:cost_homd}
\end{subfigure}
\caption{Computational cost ratio CCR and relative error in terms of the total free-energy for case 2 (a,b), uni-axial compression, and case 3 (b,d), torsional deformation, for FE size equal to 1 cm (a,c) and to 0.5 cm (b,d). CCR=100 means that the homogenized model is 100 times faster than the micromechanical one.}
\label{fig:cost_hom}
\end{figure*}

Figure \ref{fig:cost_hom} shows the CCR in terms of the relative error of the total free-energy for the most representative loading scenarios: lcC and lcT. Figures \ref{fig:cost_homb} and \ref{fig:cost_homd} correspond to the selected FE mesh size of 0.5 cm, while Figures \ref{fig:cost_homa} and \ref{fig:cost_homc} display the CCR and error for a coarser FE mesh, with size equal to 1 cm. The CCR spans between approximately 1 and 300, depending on the unit-cell dimensions. Similar trends of the CCR are observed for the other loading scenarios.

For sufficiently small values of $\epsilon$, CCR increases considerably, which shows the efficiency of the proposed method without significantly affecting the accuracy of the FE results. Notice that the FE performance can be further improved by employing adaptive mesh refinement techniques that will refine the mesh only where needed (regions of high stress gradients). This is expected to increase even more the CCR and therefore the efficiency of the calculations. Alternatively, higher order homogenization schemes and/or generalized continua may be used to improve performance due to the presence of boundary layers.

The results in this Section show the ability of TANN to deliver reliable material description of microstructured materials at reduced calculation cost. Furthermore, the homogenized model, by virtue of the thermodynamics consistency of TANN, is found to give accurate, thermodynamically consistent results. Last but not the least, the homogenized model is much less computationally demanding.

\section{Conclusions}
\noindent The bottleneck of computational homogenization methods is the solution of the auxiliary problem that emerges from the microstructure. Due to inelasticity, the solution of this problem is cumbersome, time consuming and has to be performed a huge amount of times in each numerical simulation. Here we propose an alternative method that, based on physics-aware artificial neural networks, enables to learn the nonlinear, irreversible mechanical behavior of the microstructure and then use it for performing large, multiscale numerical simulations at reduced cost. The proposed method builds on our previous work on Thermodynamics-based Artificial Neural Networks (TANN). As such, it inherits all the advantages of TANN that were extensively presented in \cite{masiTANNspigl, masi2021thermodynamics}. Consequently, our methodology provides predictions that are always thermodynamically consistent, accurate and robust.

Key point in the current work is the discovery of the necessary internal state variables from the internal coordinates of the microstructure. The internal coordinates can include the displacements, the velocities, the stresses and any other microstructural field. This identification was accomplished by using a new ANN architecture that involves an encoder and a decoder that are fed with the internal coordinates of the system. The training of this new ANN architecture is performed under the thermodynamics constraints imposed by the theory of thermodynamics with internal state variables \citep{coleman1967thermodynamics,maugin1994thermodynamicsA}. This renders possible the automatic, unsupervised identification of the necessary internal state variables that encapsulate the non-linear, inelastic evolution of the microstructure. As a result, there is no more need to heuristically define the internal state variables, which, for complex, inelastic microstructures, may be hard to identify empirically.

The framework proposed is general and can be applied in a wide range of inelastic microstructures for which computer models are available. In order to illustrate our methodology, we provide examples of two- and three-dimensional regular and non-regular lattice structures, which are made of elastoplastic elements with hardening. Once trained, the extended TANN provides accurate predictions of the macroscopic quantities such as the stress, the free-energy and the dissipation rate for a given strain increment. This was possible thanks to the identified internal state space, which can be also decoded in order to provide detailed predictions about the evolution of the microstructural fields. It is worth emphasizing that the identified internal state variables contain all the necessary information of the microstructure at a given state. Consequently, there is no need to take into account the history of any macro- or micro-scopic field as it is the case in other approaches \cite[][among others]{Wang2018,ghavamian2019accelerating,SAHA2021113452}. This is an advantage of the current method for retrieving reduced models of inelastic complex materials, which, inevitably, can show path dependency, at the macroscale, and chaotic behavior, at the microscale. Moreover, our approach allows the generation of data-sets based on random incrementation, rather than on prescribed stress paths (e.g. triaxial, biaxial, monotonous, cyclic etc.) that other methods implicitly require. This provides greater flexibility in data generation and reduces the need of huge data-sets containing rich collections of stress paths.

Based on the accurate and robust predictions of the extended TANN for the microstructure, we present examples of double-scale numerical homogenization of three-dimensional periodic lattice structures. The multiscale problem is solved using a classical, displacement based, incremental Finite Element (FE) formulation \cite{geolab}, where the material used is the trained TANN network. For the upscaling, we use a first order, asymptotic, periodic homogenization scheme, adapted to an incremental formulation \cite{miehe2002strain}, in order to account for non-linearities. The results from the Finite Element solution of the macroproblem are compared to detailed, micromechanical simulations for various sizes ($\epsilon$) of the elementary cell. The predictions of the homogenized model based on TANN show excellent agreement with the cumbersome, but exact micromechanical simulations. The predictions of the homogenized TANN model show high accuracy both at the macroscale (e.g. prediction of average stresses, strains, elastic energy and dissipation) and at the microscale (e.g. microstructural fields of displacements and forces in the lattice elements). This is confirmed for a large variety of stress-strain paths (e.g. monotonous or cyclic). Convergence with the exact solution of the problem was also observed as $\epsilon\rightarrow 0$. 

The computational cost for the solution of the homogenized problem is found to be several order of magnitudes lower than that of the detailed, exact micromechanical model. The approach can be used in any classical FE code. This enables the use of the existing artillery and know-how in FE for further accelerating the numerical calculations and increasing their accuracy (see parallelization, optimized solvers, adaptive mesh refinement etc.). Moreover, TANN are independent of the adopted up-scaling scheme (here asymptotic homogenization). Other upscaling schemes could further improve the accuracy of the macroscale calculations and capture more efficiently boundary layer phenomena, that first-order homogenization schemes cannot capture accurately. Our methodology can be also used for micromorphic/generalized continua \cite[see][]{germain1973method,forest2020continuum,GODIO2017168} by simply expanding the stress-strain space.

It is worth mentioning that the proposed methodology can cover any inelastic material. This is thanks to the general thermodynamic framework and the dimensionality reduction scheme employed. However, here we focused only on elastoplastic lattice structures, as they allow to easily present the efficiency of the method.\\
Practical applications to other materials and microstructures is the objective of future works.

Despite the use of the proposed methodology for homogenization, another application of TANN could be in constitutive modeling. The unsupervised identification of the state space can be used for understanding the involved physics of complex materials. Feature extraction techniques  \citep[see][]{strofer2018data,lu2020extracting} could be used for getting further insight into the physical meaning of the identified state space. However, this investigation exceeds the scope of the current work and will follow in the future. Finally, another application of our approach, which was not covered here, is the development of macro-elements for complex, inelastic systems \cite[e.g.][]{Grange2008,Greco2017}. In this case, TANN can be seen as a thermodynamics-based, model order reduction technique.

\section*{Acknowledgments}
\noindent The authors would like to acknowledge Dr. Alexandros Stathas for the fruitful discussions.

\noindent The authors would like to acknowledge the support of the European Research Council (ERC) under the European Union Horizon 2020 research and innovation program (Grant agreement ID 757848 CoQuake).

\section*{Data Availability}
\noindent The codes accompanying this manuscript are available at \href{https://github.com/flpmasi/TANN-multiscale}{\texttt{github.com/flpmasi/TANN-multiscale}}.

\bibliography{Bibliography}  

\section*{Appendix A} 
We present here the scheme for solving the boundary value problem of the microstructure (auxiliary problem), which is used given in Section (\ref{sec:app1}).\\
The variational form of the balance equations over the lattice in absence of bulk and inertia forces is
\begin{equation}
\int_{\mathcal{V}_{el}} \widehat{\sigma} : \delta \varepsilon\, dV - \int_{\partial {\mathcal{V}_{el}}} \widehat{t} \cdot \delta u \, dS -\int_{\partial {\mathcal{V}_{el}}} \delta \left(\widehat{\lambda} \cdot ( \widehat{u_{\Delta}^{p}} -E \cdot x_{\Delta}^p) \right)\, dS = 0,
\label{eq:int_var_form}
\end{equation}
where $\delta(.)$ denote variations, $\mathcal{V}_{el}$ the volume occupied by the lattice elements ($\mathcal{V}=\mathcal{V}_{el}\cup\mathcal{V}_{0}$, where $\mathcal{V}_{0}$ is the volume occupied by the voids), $\sigma$ is the microscopic stress tensor; $\varepsilon$ and $E$ are the microscopic and macroscopic (average) strain tensor, respectively; $t$ is the traction vector. Hat superscripts denote time incrementation (e.g. $\widehat{\sigma}\triangleq\sigma^{t+\Delta t}-\sigma^t$). 

The periodicity of the displacement field at the boundary $\partial {\mathcal{V}_{el}}$ is imposed via Lagrange multipliers $\lambda$, such that $u_{\Delta}^p= E \cdot  x_{\Delta}^p$, where $u_{\Delta}^p=u^{p+}-u^{p-}$ and $ x_{\Delta}^p=x^{p+}-x^{p-}$. $p+$ and $p-$ denote values evaluated at opposite boundaries. Due to the antiperiodicity of traction vectors and the  periodicity of the displacement, the second right-hand side term in Equation (\ref{eq:int_var_form}) is equal to zero.

By further assuming homogeneous strains over the lattice elements (bars), we obtain the following weak form of equilibrium, in terms of the nodal displacements of the elements:
\begin{equation}
\begin{cases}
&\sum_k^{N_{n}} \widehat{F}_k \cdot \delta U_k = 0,\\
&\sum_k^{N_{n}^p} \left[\delta \widehat{\lambda}_k \cdot \left(\Delta  \widehat{U}_k^{p}-E\cdot \Delta x^p \right) + \widehat{\lambda}_k\cdot \Delta  \widehat{U}_k^{p}\right]=0,
\end{cases}
\label{eq:var_form}
\end{equation}
with $F_k$ and $U_k$ being the nodal force and displacement of node $k$, respectively; $N_n$ the number of nodes of the lattice elements; and $N_n^p$ the number of nodes at the boundaries.

Equation (\ref{eq:var_form}) can be rearranged to the form $\sum_k^{N_{n}} \widehat{R}_k \cdot \delta U_k = 0$, which must hold for every $\delta U_k$, thus equilibrium is satisfied when $\widehat{R}_k=0$. The subsequent system of nonlinear equations is solved by means of the Newton-Raphson's method.

\section*{Appendix B}
\begin{figure*}[h]
\centering
\includegraphics[width=0.4\textwidth]{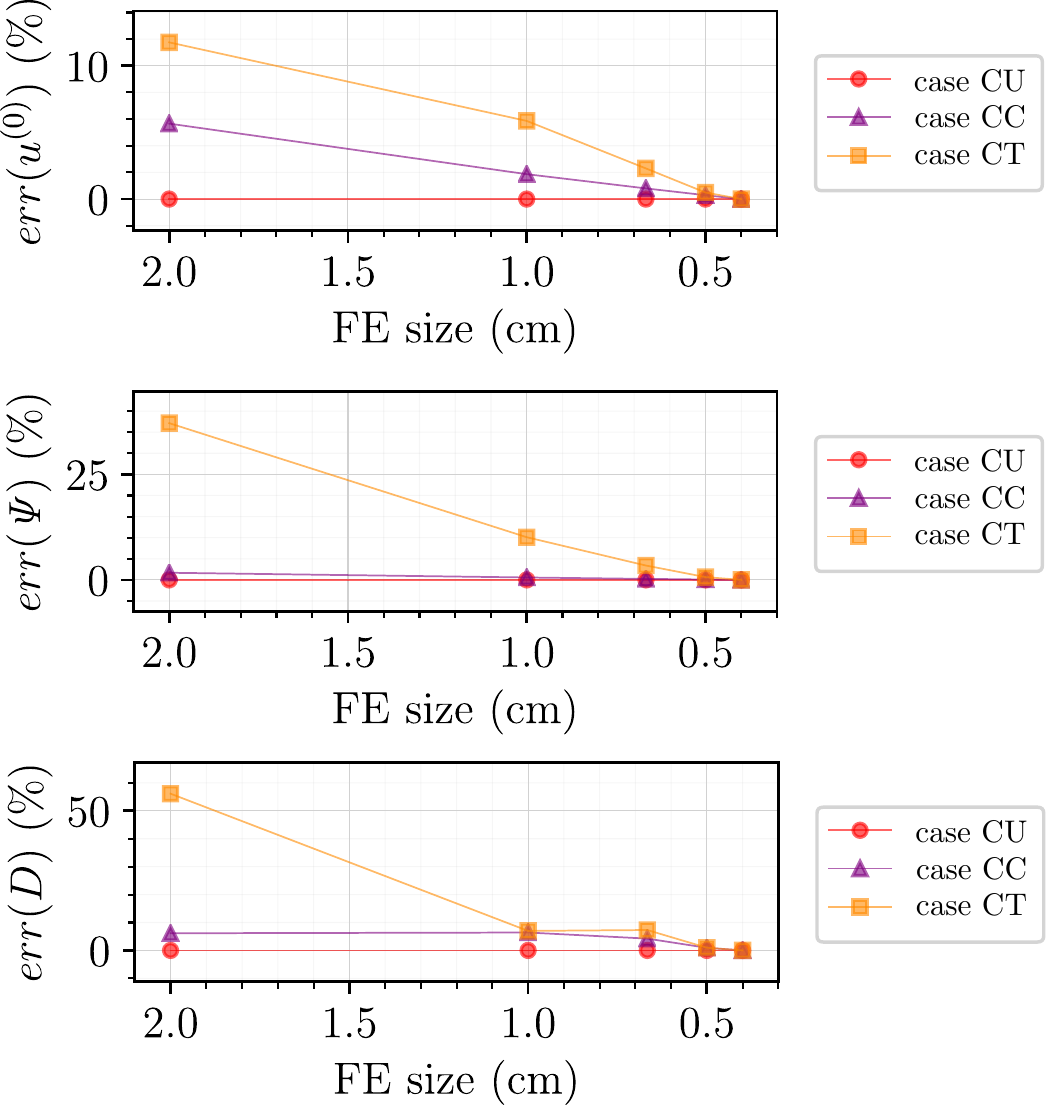}
\caption{Relative error of the displacement at the control points (top), the total energy (center), and the total dissipation rate (bottom) in function of the FE size.}
\label{fig:convergence}
\end{figure*}

We present herein the mesh sensitivity analyses performed for the Finite Element model and the numerical simulations performed in Section \ref{sec:app2}. In particular for each one of the loading scenarios lcU, lcC, and lcT, we evaluate the relative error of the zeroth-order approximation of the displacement at the control points (see Figs. \ref{fig:hom_uni}, \ref{fig:hom_comp}, and \ref{fig:hom_tors}), the total energy and dissipation rate. The errors are presented in Figure \ref{fig:convergence}. Based on these results, we choose a FE mesh with tetrahedral elements with size equal to 0.5 cm, which for all scenarios gives an relative error smaller than $1\%$, with respect to the finest mesh.

\end{document}